\documentclass{article}
 \usepackage[paper=letterpaper,dvips]{geometry}
 \usepackage{amsmath,amsfonts,url,graphicx,amscd,gastex}
 \usepackage{amssymb}
 \usepackage{enumitem}
  \usepackage{tikz}
  \tikzset{>=latex}
 \input pack10.sty 

\makeatletter
\long\def\proofbox#1{\gdef\@proofbox{#1}}
\proofbox{\small{\tt\@ifundefined{sourcepath}{}{\sourcepath/}\jobname}\\%
\ifx\UNDEF\Version\else\quad v.\Version, \fi\today}

\def\proofref#1{\proofbox{\small{\tt#1\par
	[edited by Tom Toffoli for personal use]\par
	{\tt\sourcepath/\jobname}, 
	\number\month/\number\day/\number\year
	\par}}}

 \def\affil#1{\\{\small\sl#1\par}}
 \long\def\author#1{\gdef\@author{#1}}
 \author{Tommaso Toffoli ({\tt tt\char"40bu.edu})\affil{Electrical and
Computer Engineering, Boston University, MA 02215}}

 \long\def\abstract#1{\gdef\@abstract{#1}}
 \abstract{}

\long\def\@firstoftwo#1#2{#1}
\long\def\@secondoftwo#1#2{#2}
\def\@ifundefined#1{%
  \expandafter\ifx\csname#1\endcsname\relax
    \expandafter\@firstoftwo
  \else
    \expandafter\@secondoftwo
  \fi}
 \@ifundefined{leftfrac}{\def\leftfrac{.32}}{}
 \@ifundefined{ritefrac}{\def\ritefrac{.68}}{}

\def\@maketitle{\newpage\noindent\leavevmode
  \begin{minipage}[t]{\leftfrac\textwidth}
    \hrule height0pt
    \@proofbox
  \end{minipage}\hfil
 \begin{minipage}[t]{\ritefrac\textwidth}
    \hrule height0pt
    \raggedleft
    \LARGE\@title\par
    \vskip4pt
    \large\@author
  \end{minipage}
  \vskip8pt
  \ifx\@abstract\@empty\else{\vskip.5em\leftskip1.25in\parskip4pt\small\@abstract\par\vskip.5em}\fi
  \noindent
  \rule{\textwidth}{0.4pt}
  \vskip16pt}

\makeatother

\def\leftfrac{.30}\def\ritefrac{.70}
 \proofbox{\small From {\sl Entropy \bf18}:7 (Jun 2016), 247; here reformatted
and slightly edited (May 2017)}

 \newcommand{\ie}{i.e.,}
 \newcommand{\eg}{e.g.,}
 \newcommand{\cf}{cf.~}		

 \mathchardef\BY="0202
 \newcommand{\by}{\ensuremath{\BY}}              

 \makeatletter
 \newcommand{\sectlabel}[1]{\label{sect:#1}}
 \newcommand{\footlabel}[1]{\label{foot:#1}}
 \newcommand{\eqlabel}[1]{\label{eq:#1}}
 \newcommand{\figlabel}[1]{\label{fig:#1}}
 \newcommand{\Sect}[2][]{\def\t@mp{#1}%
\section{#2} \ifx\t@mp\@empty\else\sectlabel{#1}\fi}
 \newcommand{\Subsect}[2][]{\def\t@mp{#1}%
\subsection{#2} \ifx\t@mp\@empty\else\sectlabel{#1}\fi}
 \newcommand{\Foot}[2][]{\def\t@mp{#1}%
\footnote{#2\ifx\t@mp\@empty\else\footlabel{#1}\fi}} 
 \newcommand{\Eq}[2][]{\def\t@mp{#1}%
\begin{equation}#2\ifx\t@mp\@empty\notag\else\eqlabel{#1}\fi\end{equation}}
 \newcommand{\Eqaligned}[2][]{\def\t@mp{#1}%
\begin{equation}\begin{aligned}#2\end{aligned}
\ifx\t@mp\@empty\notag\else\eqlabel{#1}\fi
\end{equation}}
 \newcommand{\sect}[1]{\S\ref{sect:#1}}		
 \newcommand{\foot}[1]{Footnote~\ref{foot:#1}}	
 \newcommand{\eq}[1]{(\ref{eq:#1})}		
 \newcommand{\fig}[1]{Fig.\,\ref{fig:#1}}

\newcommand{\Fig}[3][]{
\begin{figure}[!htb]
 \centering{\leavevmode#2}%
 \caption{#3}
 \figlabel{#1}
\end{figure}                 }
 \makeatother	

 \def\squeeze{\itemsep0pt\parsep0pt\parskip2pt} 
 \def\pages#1{}

 \title{Entropy? Honest!}				
 \def\ABSTRACT{Here we deconstruct, and then in a reasoned way reconstruct, the
concept of ``entropy of a system,'' paying particular attention to where the
\emph{randomness} may be coming from. We start with the core concept of entropy
as a \emph{count} associated with a \emph{description}; this count
(traditionally expressed in logarithmic form for a number of good reasons) is
in essence the \emph{number} of possibilities---specific instances or
``scenarios,'' that \emph{match} that description. Very natural (and virtually
inescapable) generalizations of the idea of \emph{description} are the
\emph{probability distribution} and of its quantum mechanical counterpart, the
\emph{density operator}.

We track the process of dynamically \emph{updating} entropy as a system
evolves. Three factors may cause entropy to change: (1) the system's
\emph{internal dynamics}; (2) unsolicited \emph{external influences} on it; and
(3) the approximations one has to make when one tries to predict the system's
future state. The latter task is usually hampered by hard-to-quantify aspects
of the original description, limited data storage and processing resource, and
possibly algorithmic inadequacy. Factors 2 and 3 introduce
\emph{randomness}---often huge amounts of it---into one's predictions and
accordingly degrade them. When forecasting, as long as the entropy bookkeping
is conducted in an \emph{honest} fashion, this degradation will \emph{always}
lead to an entropy \emph{increase}.

To clarify the above point we introduce the notion of \emph{honest entropy},
which coalesces much of what is of course already done, often tacitly, in
responsible entropy-bookkeping practice. This notion---we believe---will help
to fill an expressivity gap in scientific discourse. With its help, we shall
prove that \emph{any} dynamical system---not just our physical
universe---strictly obeys Clausius's original formulation of the second law of
thermodynamics \emph{if and only if} it is invertible. Thus this law is a
\emph{tautological property} of invertible systems!}

 \abstract{\ABSTRACT}

\newcommand{\spade}{\ensuremath\spadesuit}
\newcommand{\club}{\ensuremath\clubsuit}
 \long\def\xtra#1{\smallskip\noindent{\footnotesize#1\par}\smallskip}

\makeatletter
\newcommand{\myquote}[2][]{\def\t@mp{#1}\hfill\begin{minipage}{3in}\sf\small#2%
\ifx\t@mp\@empty\else\par\smallskip\noindent---\t@mp\fi%
\end{minipage}\bigskip}
\newenvironment{dialog}{\par}{\par}
 \def\dir#1{(\emph{#1})}
 \newcommand{\cue}[3][]{\def\t@mp{#1}\par\noindent\hangindent1.6em\hangafter1{\sc#2}\ifx\t@mp\@empty\else\;[\t@mp]\fi:\hskip.75em#3}

 \def\lawidth{3.1in}
 \newcommand{\Law}[2][]{\def\t@mp{#1}%
\begin{equation}\parbox{\lawidth}{\centering{\sc#2}}\ifx\t@mp\@empty\notag\else\eqlabel{#1}\fi\end{equation}}
 \makeatother

\begin{document}
\maketitle

\myquote[{\sl Entropy}'s Referee 1]{This is a great paper. I worked on
thermodynamics for many years, and entropy is at the centre of my research in
this area, yet this paper made me see that there were many things I thought I
understood but actually didn't, and clarified many of them for me. \dots The
style is unique. I honestly can say that I never read anything like this and
that I enjoyed it thoroughly. The last thing I would like is for the style to
be changed, even though it is far from that of a standard scientific
paper.}

\myquote[{\sl Entropy}'s Referee 2]{This paper is a delight.  I have studied
and written about entropy for years, and find the definition [of entropy] as
``a function that assigns a number to a \emph{description}'' really hits the
nail on the head.  This paper is original and significant; \dots in sum,
marvelous paper.}

\Sect[intro]{Introduction}

\myquote[Paraphrase of Luke 23:34]{Forgive them, for they mean not what they
say.}

In recent decades, entropy has come to be applied very broadly.  And like a few
other scientific constructs which had ``made the grade'' in earlier
decades---such as evolution, relativity, and quantum mechanics---entropy too
has found favor with the general public and somehow managed to become a
household word.  It carries the right combination of glamor, tantalizing
promise, and prurient mystery. It comes in handy in general conversation,
without requiring of the parties much commitment or understanding. This of
course abets vagueness, confusion, misuse, and abuse, and the propagation of
``urban legends'' about it that occasionally penetrate even responsible
scientific quarters (see the \emph{Shannon} entry in \sect{entropy000}).

My primary purpose here is to review how and why the entropy ``of a system''
(the scare quotes, as we shall see, are deliberate) evolves in time as the
system itself evolves; and to stress of what a radically different \emph{nature}
those two evolution processes are. To anticipate, the second process is in the
nature of \emph{bookkeeping}, which, to be useful, must first of all be
\emph{honest}.

As part of this endeavor, I will motivate and then introduce the concept of
\emph{honest entropy}. This fills, I believe, a gap in the literature of
entropy and the second law of thermodynamics.

\medskip

To help concentrate minds, my rhetorical (and pedagogical) device will take the
form of a campaign to dispel a number of widespread \emph{myths} about entropy,
such as:
 \begin{enumerate}\squeeze
  \item The belief that, in spite of having the same name and sporting the same
formula, \emph{information-theoretical} entropy (\`a la  Shannon and Jaynes) is
actually ``something completely different'' from \emph{physical} entropy (\`a la
Clausius, Boltzmann, and Gibbs).
  \item The myth---\emph{pace} Clausius---that physical entropy is a property
of a definite material \emph{body} one can point to---such as the copper bar
now lying on my lab bench---and is a \emph{physical} property, like volume,
temperature, composition, internal energy, etc., that anyone can objectively
determine by physical means and track as it changes in time.
  \item The myth that the approach of an isolated system toward its maximum
entropy, though generally \emph{monotonic} (\ie without ever changing
direction) as predicated by Clausius, is actually subject to \emph{statistical
fluctuations} \`a la Ehrenfest (see \sect{ehren})---with entropy occasionally
\emph{decreasing} rather than increasing.
  \item The myth that the (information-theoretical) entropy of the deck of
cards now lying on my table is an \emph{intrinsic} property of it, which anyone
can determine by inspecting the deck itself.
  \item The belief that \emph{logarithms}---as encountered in Boltzmann's
formula $S=k\log W$ and Gibbs's and Shannon's $S=-k\sum_i
p_i\log p_i$ are an \emph{essential} feature of the entropy concept.
 \end{enumerate}
 To add insult to injury, I'll throw in two more old chestnuts:
 \begin{enumerate}[resume]\squeeze
 \item The myth that entropy, whether physical or information-theoretical, is
something that you \emph{measure}.
 \item The myth that the second law of thermodynamics is a law of
\emph{physics} (please wait until \sect{entropy001} before you cast your stone).
 \end{enumerate}

 \Sect[myths]{Myths: pros and cons}

 Relax now!
 I wholeheartedly grant that all good myths have at their core an element of
truth, and that their telling and retelling may have pedagogical value. I
myself can think of many circumstances where any one of the above myths may
serve a useful purpose---as an aid to intuition, in popular science; as a
stepping stone, in teaching; and in ordinary scientific discourse, as a
convenient abbreviation when the underlying core truth is understood. Let's
take the following dialogue as a case study.
 
 \def\Te{Teacher}
 \def\Pu{Pupil}

 \medskip
 \begin{dialog}
 \cue{\Te}{Think of random number between 1 and 10.}
 \cue{\Pu}{Seven!}
 \end{dialog}

 \xtra{As an aside on the limitations of human psychology, let me remark
that the empirical odds are about 1:1 in favor of this answer.}

 \begin{dialog}
 \cue{\Te}{What? \dir{sarcastically} What may make you think that \emph{seven},
of all numbers, should be random? It's an ordinary, perfectly definite number,
like 1, 2, and 3---in fact, everyone knows it comes right after 6!}
 \cue{\Pu}{\dir{blushes, embarrassed}}
 \cue{\Te}{Class, {\sc Home assignment}: List all numbers from 1 to 100 and put
a mark next to those that are \emph{odd}. On a separate column, mark all those
that are \emph{prime}. On a third column, mark all those that are
\emph{random}!}
 \end{dialog}
 \medskip

Of course, there is no such thing as a ``random number''---that is a mythical
beast.  ``Being random'' is not the property of any particular
\emph{number}---this qualifier applies to a \emph{procedure} that will generate
a number. Thus, ``Give me a \emph{random number}'' can only be understood as
shorthand for ``Use a \emph{random} procedure to generate a \emph{number}, and
give me the result.''

 \xtra{To what extent a ``random procedure'' may itself be some sort of mythical
beast, that's another kettle of fish. Even though the art has made enormous
strides since then, one should still keep in mind von Neumann's quip, that
``Anyone who attempts to generate random numbers by deterministic means is, of
course, living in a state of sin.'' (But see end of \sect{many}.)}

One may argue with John White\cite{White71} that ``Too much specific
information presented \emph{too soon} may well be aesthetically undesirable. On
the other hand, one may ask, with Gerald Dworkin\cite{Dworkin15}, ``Are these
ten lies justified?''

A compromise was acknowledged by the medieval Jewish philosopher
Maimonides. Joseph Stern\cite{Stern13} seems to conclude that the latter
actually not only acknowleges but indeed endorses, and even \emph{prescribes},
the double standard of ``[mythological] religion for the \emph{masses} and
philosophy for the \emph{elite}.''

For my part, I prefer to ask, \emph{cui prodest?} (``who has more to gain?''),
the teacher or the pupil, and on what time scale?  A short-term prop that is
cheap for the teacher today may well leave a pupil happy \emph{now} (less mental
work), but with a model of the world that may have to be gutted and rebuilt
\emph{later} (if ever)---possibly at much greater cost.

 \xtra{The ten dollars' worth of salt we spread on the driveway to save us the
pain of snow shoveling will cost the community \$200 for consequent damage to
roads, bridges, cars, water supply, and wild life.}

\noindent What I recommend instead is to help build a world model that is
\emph{honest from the start}, while only making use---at least initially---of
those constructs that are \emph{absolutely vital}. The latter is a feature, not
a regrettable concession!

\Sect[approach]{Approach}

\myquote[journalist Ralph Carpenter, 2006]{[It] ain't over till the fat lady sings!}

In that spirit, I will start by introducing a very lean concept of entropy
accompanied by a very lean concept of dynamical system, and in a few steps
arrive at a very lean-and-mean second law of thermodynamics that does all the
right things.

After that, beginning with the lean concept and making a single, modest
generalization, I'll arrive at a fully fleshed-out entropy---essentially the
one you would be familiar with. This generalization---essentially, replacing a
set by a \emph{multiset}---is pre-announced in a warning box at the beginning
of \sect{entropy001}, but is actually carried out only in section
\sect{decon}.

I'll finally argue that such generalization is not even strictly necessary (and
possibly not even sufficient).  It's just a convenient shortcut, and besides
it's not universally applicable.  One could get along just as well, if not
better, by using a more sophisticated scenario as a ``wrapper'' for the lean
entropy we had started with. This approach (\sect{many}) may not be practical,
since it doesn't carry out any simplifications until the very last step (\cf
the present section's epigraph), and is thus exceedingly cumbersome---but it's
a good device for thought experiments.

\medskip

All of the above is directly applicable not only to physics and communication
channels, but also to scenarios of other sciences as well as to \emph{much of
ordinary life}.

\bigskip

On reading the above, many a one of you will respond with a dismissive sneer, ``I
already knew all of that!''
 and, pointing to a twelve-foot stack of books and papers on their office floor
labeled {\sc entropy and the second law}, add ``In fact, I have it all
here!---ehem \dots\ someplace around here.''

Others, I hope, will go to \emph{their} offices, and beaming with relief will
throw out most of \emph{their} twelve-foot stacks, keeping only a few trusted
pieces---thus making room (in their brains as well as their offices) for the
new, exciting stuff that I'm sure will later show up in their lives.  Pointing
to the small stack left they'll say with pride, after Hillel, ``This is the
entire thing---\emph{all the rest is commentary}!''

 \xtra{When a potential convert approached Babylonian rabbi Hillel and asked
whether he could be taught the entire Torah while standing on one foot---the
1st century {\sc bc} equivalent of an ``elevator speech''---Hillel summarized
as follows, ``That which is hateful to you do not do to others. All the rest is
commentary. Now go and learn.''[Shabbos 31A]}

\noindent They will surely add with bated voice (as I imagine Hillel did too)
``And, for that matter, mostly frivolous, repetitious, superfluous, of merely
archival nature, or too technical---when not actually \emph{wrong}!''

\medskip

Familiarity with the concept of entropy is useful in all walks of life. Entropy
rightly belongs to \emph{everybody}---kind of a ``{\sc Unesco} World Heritage
Concept.'' But how many are in a position to step up and claim their
inheritance? Not every Tom, Joan, and Harry can afford the space, in their
brains as well as in their living rooms, for a twelve-foot stack of papers---or
have the time, the preparation, and the discipline to wade through them!

Let's face it, outside of our specialty we are all Toms, Joans, and Harrises,
and the baggage of our everyday interests and projects leaves little space
in our lives for extra clutter. \emph{Pace} Maimonides, I have attempted a
drastic compression of the resources needed for \emph{everybody}---not only the
specialist---to understand ``entropy and all that'' and make an honest use of
it.  Compressing by popularizing, mythologizing, and patronizing is easy, but
tends to throw out the baby with the bathwater; I wanted mine to be a
\emph{lossless} compression.

Working on this project was great fun, especially thinking of how much room it
would make \emph{in my own life} for new projects.

\Sect[entropy000]{Entropy 000}

 \def\namedate#1#2{\noindent{\bf#1} (#2)\par} 

Before attempting a reconstruction of entropy, let us see what kind of edifice
we have to begin with, whose functionality we aim to duplicate by the most
frugal means. Here is a brief historical sketch.

\bigskip\namedate{Clausius}{1850} For German physicist Rudolf Clausius, entropy
is a \emph{physical quantity} pertaining to a \emph{body}, like internal energy
and mass. It has the physical dimension of \emph{heat capacity}, denoted by
$\text{\sf E}\text{\sf T}^{-1}$, where {\sf E} means a quantity ``of the
\emph{energy} type'' and {\sf T}, ``of the \emph{temperature} type.''
Accordingly, it's measured today in units of \emph{joules} per \emph{kelvin}
(J/K).
 
Clausius wrote nine memoirs between 1850 and 1865, collected with an
introduction and appendices in \cite{Clausius65heat}. The last few lines of the
ninth memory say
 \begin{quote} [Having introduced \emph{entropy} beside energy,] we may express
in the following manner the fundamental laws of the universe which correspond
to the two fundamental theorems of the mechanics of heat:
 \begin{center}
 1. {\sc The energy of the universe\\ is constant.}\\
 2. {\sc The entropy of the universe\\ tends to a maximum.}
 \end{center}
\end{quote}
 
\noindent Those are the famous two \emph{laws of thermodynamics}.\Foot
 {For the casual reader, a popular but scientifically solid introduction may be
found in \cite{Kirwan00}.}
 They sport a poetic concision comparable to the last few
lines of Darwin's \emph{Origin of Species} (appeared at about the same time, 1859), which say
 \begin{quote} Whilst this planet has gone cycling on according to the fixed
law of gravity, from so simple a beginning endless forms most beautiful and
most wonderful have been, and are being, evolved.
 \end{quote}

\medskip

\xtra{The quantity that Clausius thought up would tell
under what conditions a thermodynamical ``body'' (or ``system,'' to use a more
modern term) could progress from a given initial state to a desired final state
\emph{spontaneously}---that is, without being pushed by an external
agency. Clausius had initially called this quantity (for which he proposed the
symbol $S$) the ``transformation content'' of the body, but later himself
coined for it the term ``entropy,'' in analogy with physical ``energy'' (having
especially in mind the body's internal energy $U$, comprising both the
mechanical and the thermal energy stored in the body itself).  In his words,

 \begin{quotation}
 \noindent We might call $S$ the transformation content of the body, just as we
termed the magnitude $U$ its thermal and ergonal content. But as I hold it that
terms for important magnitudes had better be made up from the ancient
languages, so that they may be adopted unchanged in all modern languages, I
propose to call the magnitude $S$ the \emph{entropy} of the body, from the
Greek word \emph{trop\=e}, transformation. I have intentionally formed the word
entropy so as to be as similar as possible to the word energy, for the two
magnitudes to be denoted by these words are so nearly allied in their physical
meanings that a certain similarity in designation appears to be
desirable\cite[9th memoir]{Clausius65heat}.
 \end{quotation}

\noindent The wordplay here is that in Greek \emph{trop\=e}
and \emph{\'ergon} respectively mean `transformation' and `work' (as well as
`activity', `energy', `force'), and \emph{en} means `in', so that `en-ergy' can
be construed as ``work-in-it'' or ``work contents'' (note that the German for
`contents' is \emph{In-halt} or `in-held'); by the same token, `en-tropy' may
well be taken to literally mean ``transformation contents.''

}

\bigskip\namedate{Boltzmann}{1877} Boltzmann's out-of-the-box intuition was that
entropy is not a material quantity (``Five gallons of entropy---and check the
tires!''), but just the logarithm of a \emph{count}. In his formula
 \Eq{
	S = k \ln W
  }
  $S$ denotes entropy, $\ln$ is the natural logarithm (\ie log in base $e$),
and $W$ the number of \emph{complexions}---as explained in a moment. The
coefficient $k$ appends---to what is otherwise a \emph{pure number}---the
physical dimensional unit required to remain consistent with Clausius's usage.

Originally, those complexions were imagined to be the different ways that
thermal energy can rearrange itself within a body of a certain
description. These fine-grained internal arrangements roughly correspond to what
today we call a body's \emph{microstates} (while a \emph{macrostate} is the
very description of the body in terms of the {whole set} of microstates it can
possibly be found in). Note that Boltzmann's ``body'' is already a more abstract
entity than Clausius's.

 \xtra{For Boltzmann, a ``body'' is no longer that one scrawny
plucked chicken that you brought back from the farmer's market, but an entry in
the supermarket's flyer---a \emph{type} of product. That's why when you meet your
neighbor at the supermarket you can tell her, ``I bought the \emph{same}
chicken as yours!'' even though you two walk home with a pullet apiece.

Similarly, when the four customers from table 7 all ordered the \emph{same}
chicken dish  they cannot complain if the four servings were not literally ``the
same''---there certainly were some small differences. In \emph{how many ways} a
serving can qualify as being a match for a given menu item, this may be just a
philosophical question for a chef; but it may lead a physicist like Boltzmann to a revealing insight.}

\bigskip\namedate{Gibbs}{1902} Gibbs's idea of entropy is captured by the formula
 \Eq[gibbs]{
	S=-k\sum_i p_i\ln p_i,
 }
 in current use today.

 \xtra{Note that it is Gibbs that coined the name of the whole discipline,
\emph{statistical mechanics}. Incidentally, Albert Einstein was unaware of
Gibbs's contributions in that field when, between 1902 and 1904, he wrote three
papers on statistical mechanics.  After reading Gibbs's textbook, in the 1905
German translation by Ernst Zermelo, Einstein declared that Gibbs's
treatment was superior to his own and explained that he would not have written
those papers if he had known Gibbs's work\cite{Navarro98}.}

Gibbs's formulation\cite{Gibbs02} represents a modest generalization of
Boltzmann's approach---not a radical upheaval. Instead of individual
microstates, the $i$'s in the above formula represent \emph{kinds} of like
microstates, and $p_i$ denotes what \emph{share} of a certain population is
represented by individuals of kind $i$. In other words, instead of explicitly
counting \emph{individual} microstates Gibbs counts \emph{kinds} of
microstates, but weights them in terms of \emph{what fraction} of the total
number of microstates each kind represents.

\medskip

{\bf Note:} This is the generalization mentioned at the beginning of
\sect{approach}.

\bigskip\namedate{von Neumann}{1927}

\xtra{In quantum mechanics, the concept of macroscopic state of a system is
captured by a \emph{density operator}\cite{vonNeumann27,Petz01} (co-discovered
by Lev Landau), a composite of an \emph{empirical uncertainty} already found in
classical mechanics (``in principle we could always get a sharper picture of
the state of a system through a more refined observation'') and an
\emph{irreducible uncertainty} peculiar to quantum mechanics (``there are
aspects of a system that we can know more about only at the cost of ending up
knowing less about other aspects of the same system).}

Von Neumann's entropy of a \emph{density operator}---which is a generalization
of a probability distribution---is defined as
 \Eq[vonneumann]{
	S=-\kappa\sum_i \rho_i\ln \rho_i,
 }
 where the $\rho_i$'s are the \emph{eigenvalues} of the operator. Formally this
is much like Gibbs's expression.

According to standard quantum mechanics, there are concretely testable aspects
of a physical system that cannot be accounted for by simply postulating a
description of the whole universe in terms of a single \emph{overarching}
probability distribution over it (this is an implication of \emph{Bell's
inequality}), but where, as far as physicists know, a density operator will do
(\cf \sect{many}).

\bigskip\namedate{Shannon}{1948} To answer questions of how much information
could be ``squeezed'' through an information channel such as a telegraph line,
Shannon came out in 1948 with a fully fleshed-out \emph{Information
Theory}\cite{Shannon48,Lesne14}. In this theory, the expression for entropy is
 \Eq[shannon]{
  S=-\sum_i p_i\ln p_i.
 }
 Note that this formula does away with Boltzmann's and Gibbs's ``legacy''
dimensional factor $k$ (or von Neumann's $\kappa$), so that here $S$ remains a
\emph{dimensionless} quantity---a ``pure number.''

\xtra{In April 1961 Shannon was in residence at MIT for a week, and Prof
Tribus, a pioneer at revealing the connections between information theory and
thermodynamics and at adopting Jaynes's \emph{MaxEnt} principle (see
\emph{Jaynes} entry right below), had occasions to spend time with him. He
asked Shannon whether he hadn't been afraid, when he named his
information-theoretic quantity ``entropy,'' that this would create confusion
with the original use of the word in thermodynamics. According to Tribus's
recollection's\cite{Tribus71}, Shannon replied:

 \begin{quotation}\noindent My greatest concern was what to call it. I thought of
calling it `information', but the word was overly used, so I decided to call it
`uncertainty'. When I discussed it with John von Neumann, he had a better
idea. Von Neumann told me, ``You should call it entropy, for two reasons. In
the first place your uncertainty function has been used in statistical
mechanics under that name. In the second place, and more importantly, \emph{no
one knows what entropy really is, so in a debate you will always have the
advantage!}'' 
 \end{quotation}

\noindent The emphasis is mine. Note how, a good century after the invention of
entropy, von Neumann---a polymath physicist---was well aware of how even
professional physicists still stood somewhat confused as to what entropy really
is.

Here I cannot resist a bit of gossip. In the same week, Tribus gave an MIT
seminar on a new way, based on information theory, to derive thermodynamics. He
states that a critical audience, comprised of students of American mechanical
engineer Joseph Keenan (founder of the MIT school of thermodynamics) ``tried to
rip it apart.'' Moreover, French mathematician Benoit Mandelbrot, who was in
the audience, quickly attacked the MaxEnt interpretation, saying: ``Everyone
knows that Shannon's derivation is in error.''\cite{Tribus61}

That shows what strong feelings people may have about ``mere matters of
interpretation'' of entropy and thermodynamics.}

\bigskip\namedate{Jaynes}{1952} ET Jaynes was an ardent developer and preacher
of the Bayesian interpretation of statistical mechanics.  He ``emphasized a
natural correspondence between statistical mechanics and information theory. In
particular, [he] offered a \emph{new and very general rationale} why the
Gibbsian method of statistical mechanics works. He argued that the entropy of
statistical mechanics and the information entropy of information theory are
\emph{principally the same thing}. Consequently, statistical mechanics should
be seen just as a particular application of a general tool of \emph{logical
inference}'' (\cite{Wikipedia16Jaynes}; emphasis mine).

He's best known for the introduction of the \emph{principle of maximum entropy}
(or \emph{MaxEnt})\cite{Jaynes57maxent,Hanel14}. Adapting from \cite{Toffoli04},
 \begin{quotation}
 \noindent If, in a given context, you need to formulate a probability
distribution on which to base your bets, choose, among all possible
distributions that agree with what you know about the problem, the one having
\emph{maximum entropy}. Why? Is this guaranteed to be the ``real'' (whatever
that may mean) probability distribution?  Of course not! In fact you will most
likely replace it with a new one as soon as you see the outcome of the next
trial---because by then you'll have one more piece of information. Why, then?
Because any other choice---being tantamount to \emph{throwing away} some of the
information you have or \emph{assuming} information you \emph{don't}
have---would be indefensible.
 \end{quotation}

\noindent This principle, then, introduces a criterion of honesty into the
otherwise poorly constrained process of inference: ideal witnesses are expected
not to just ``tell the truth,'' but tell \emph{the whole truth} (as far as each
can know it) and \emph{nothing but the truth}.

\smallskip

The concept of honest entropy presented in this paper develops and refines
the qualifiers ``as far as'' and ``nothing but.''

 \bigskip\namedate{Other entrants}{1945--}
 A number of other concepts were developed in the past seventy years in
attempts to quantify properties such as \emph{variety}, \emph{diversity},
\emph{multiplicity}, \emph{richness}, and \emph{distinctiveness}, making use of
different kinds of \emph{mean} and of \emph{scaling function}. We will mention
some of these in our deconstruction of Gibbs's formula in \sect{decon}.

\Sect[entropy001]{Entropy 001}

 \myquote[William Briggs, \url{wmbriggs.com/post/17974} (12 Feb 2016)]{If you
can't describe it, you can't put it into your equations.}

 \myquote[Luke 12:7]{Indeed, the very hairs of your head are all numbered.}

In the mercantile world, there are obvious difficulties in determining a fair
barter rate for goods of a different nature---``How much of my `land' is your
`oil' worth?''. Physics' ``merchants'' have found it expedient to value certain
of their wares in terms of a notional currency called \emph{entropy}. But
language and customs, as we know, vary from region to region and from time to
time.  Are we sure that Trader Gibbs's entropy is the same thing as Master
Boltzmann's? How come Prof Clausius's entropy grows monotonically towards a
maximum, while that used by Herr Ehrenfest may occasionally \emph{dip}? Which
of these two merchants would you rather do business with? As we shall see
(\sect{ehren}), Herr Ehrenfest must have learned to play poker at a place where
``peeking'' was allowed!

Here we shall introduce a ``primitive conceptual gage'' for entropy, a
reference benchmark against which more sophisticated entropy constructs can be
tallied.

\xtra{In biology, this role used to be played by the \emph{exemplar specimen}
of a species, kept in a museum. If there was a doubt about a naturalist's
correct identification of \emph{his} specimen, in principle one could compare
it to the exemplar and see whether (in spite of being hairier, larger, or of a
slightly different color) it displayed certain \emph{essential} characteristics
of the latter (four legs, stubby tail, suckles offspring, yet \emph{lays
eggs}---remember the ornithorhyncus?).}

\bigskip 

 As we shall see, a distinguishing trait of entropy is being of the nature of a
\emph{count}---the size (or \emph{cardinality}) of a set. ``If it ain't a count,
then it ain't entropy!''

 \bigskip
 \noindent
 \fbox{\begin{minipage}{.97\columnwidth}
 Those of you who are tempted to immediately cry ``Foul! This is not the
entropy we know!'' should wait for \sect{equi}, \sect{prob}, and \sect{many},
where probability distributions will be introduced and discussed. Similarly,
those who expect to see logarithmic scaling should wait for \sect{log}.  The
``lean'' concept of entropy---that is, entropy as a count---announced in
\sect{approach}, introduced in the present section, and used throughout for the
first two-thirds of this paper---will be complemented in the last third
(sections \sect{decon} through \sect{conclusions}---by the ``full-bodied''
concept that you would be familiar with---namely, entropy as a \emph{log} of an
\emph{equivalent} count.
 \end{minipage}
 }

 \bigskip

 What \emph{kind of bench} should provide the benchmarks to tally entropy
against? What makes a quantity ``entropy-like'' rather than, say,
``velocity-like?" When we hear ``the entropy of a \emph{body}'' (Clausius), or,
in more modern terms, ``of a \emph{system},'' our first reflex should be ``The
entropy of precisely \emph{what}? What \emph{kind} of thing can entropy be a
property of?'' We already saw that in the phrase ``prime number'' the term
``prime'' is used to qualify a \emph{number} as being prime, while in the
locution ``random number'' what is really being qualified as random is not a
number but a \emph{method} for generating one.

Let us then announce, without further ado, that
 \Law[entropy]{entropy is a property of a \underline{description},\\namely,
 the \underline{number} of items\\ that \underline{match that description}.
 }
 \noindent Thus, entropy is a \emph{function} that assigns a \emph{number} to a
\emph{description}.

\xtra{I've taken the liberty, for the sake of plainer English, to call
``description'' what is, more generally, a \emph{characterization}, that is,
something that can be used to narrow a choice, to tell \emph{by any means
available} whether or not an item belongs to a set. Take the phrase ``Be
considerate---of the prom photos in that box only pick one of which there are
more than one copy left!''  This doesn't ``describe'' a photo in the narrow
sense of ``telling what it \emph{looks like}.''  Yet in the given
circumstances, where I can see all the photos in the box at once, it surely
characterizes a definite subset of them.}

Having clarified the meaning of description, one may now ask, ``Description of
\emph{what}? Well, of practically anything of which many possibilities or
variants are available, out of which the description itself will identify a
subset. The ordering of a deck of cards; a card hand, such as
``four-of-a-kind;'' the set of possible microscopic inner arrangements of a
human-scale object; a liter of gas described as ``at standard conditions of
pressure and temperature,'' and thus entailing a certain (astronomical) number
of possible complexions of position-and-velocity of its molecules; the state of
a finite-state automaton; a generic ``winning chess move,'' as contrasted to a
specific chess move which happens to be a winning one; a computer procedure
which you call on the basis of its API description,\Foot
 {An \emph{Application Programming Interface} is a repertoire of accessible
programming commands that hide low-level (``technical'') details not needed by
the high-level programmer. In this way, application developers are free to
change ``under the hood'' details of the technical implementation without
disrupting the application's behavior. Thus an API stands for a whole
\emph{equivalence class} of low-level programs yielding the same high-level
behavior.}
 trusting that whatever code is underneath will do what the description seems
to imply; an email message waiting for you to read it; the vehicle that hit me;
whatever parasite is responsible for my illness; an ``alkaline metal;'' a
tic-tac-toe game; the next president of the United States; a plausible
universe; even \emph{another description}!

\bigskip

Here is an example of descriptions of an object.  A policeman spots a man lying
by the roadside. ``Are you hurt? What's the story?'' ``A glancing blow by a
hit-and-run vehicle, officer.'' ``That's not much help; there are perhaps a
hundred thousand vehicles within fifty miles of here---I can't stop everyone.
But, let's see, what kind of vehicle? A truck?''  ``No, it was a small car.''
``That's better---we've restricted the choice to about ten thousand cars. Did
you notice the color?'' ``Yes, it was red!''  ``Splendid---we are down to a
thousand. And when did this happen?'' ``Oh, no more than five minutes ago.''
``Then they can only have gone a few miles. Within that reach, we are talking
about a dozen cars. I will give it a try!''  (\emph{speaking on his radio})
``To all police cars: Hit-and-run accident near Springbrook's North Exit. Stop
all small red cars within five miles of there and report to me.''

\smallskip

Our target is undoubtedly \emph{one specific object}---the car involved in the
accident.  It is generally believed, since the time of Laplace, that at a
microscopic level of detail our physical world is strictly
\emph{invertible}. In principle, if at the present moment we took a
\emph{sufficiently detailed snapshot} of a \emph{large enough portion} of the
world around the site of the accident, and ran this system \emph{backwards} in
time by means of a sufficiently detailed simulation, in five minutes (of
simulated time) a usable record of the accident---car, licence plate, driver,
collision details---would materialize under our eyes.

But we can't do that \emph{in practice}---today at least!

\xtra{Though we learn fast. Today we can routinely tell \emph{paternity}, and
have started deciphering affidavits written on gravitational waves a billion
years ago and billions of billions of miles away.}

\noindent To try to bring the driver to justice, all we can use to identify the
car by is whatever circumstantial evidence we can summon up---in plain words, a
\emph{description} of the car by witnesses (even inanimate ones, such as tire
tracks). It will be up to courts and lawyers to pass in review all the
conceivable objects that \emph{match that description}, and see to what extent
each of these objects fits in with other external factual data.  For example,
if it turns out that---\emph{at the time of the accident}---a car otherwise
matching the description had been captured by a videocamera filling up at a gas
station, that particular car may be written off.

\smallskip

In the estimate of the officer, the first description (``a hit-and-run
vehicle'') gave a hundred thousand matches. A more refined one (``a small
car'') yielded one-tenth as many. The next refinement (``a red one'') further
shrank the number of matches down to one thousand. Finally, taking into account
also the timing (``less than five minutes ago''), the number of possible
choices dwindled to a dozen.

So we have four different descriptions of the same target, and each comes with
a number attached to it---the number of vehicles that match that
description. This number is not, in any sense, an intrinsic property of the
\emph{vehicle}---it is a property of a \emph{description} of it, including
external details such as where and when.

\medskip

I want to stress again that it is such a \emph{number}, associated with a
\emph{description} and reflecting \emph{how many items} fit that description,
that constitutes the essence of the \emph{entropy} concept.  In \sect{prob},
after introducing the idea of \emph{probability}---again in a basic form\Foot
 {That is, without going into measure theory, whose whole worry is about
problems caused by \emph{infinities}.}%
 ---I'll show that a very natural (and virtually inescapable) generalization of
the concept of \emph{description} given here is that of a \emph{probability
distribution}. In fact, it is on that generalization that the canonical
definition of entropy rests---the entropy ``of a probability distribution.''
But to make sense of this we'll have to clarify ``probability'' first.

\medskip

 \def\leaderfill{\leaders\hrule\hfill}

 \hbox to\hsize{{\sc A concrete example} \leaders\hrule\hfill}

 \smallskip
A colleague points at two identical-looking decks of cards on
the table, both showing as top card an Ace of Spades. ``These are ordinary
52-card decks. The first came directly from the factory---I just carefully
removed the cellophane wrapper. As for the second, I shuffled it many times,
and then continued shuffling until I got that ace back on top. Can you tell me
the entropy of the two decks?'' I reply, ``The entropy of the first deck is 1,
since only the default ordering---Ace of Spades, 2, 3, \dots, Queen, King of
Spades, and so forth for the other three suits---matches your description.  The
entropy of the second is of course 51 factorial.  In fact, the top card clearly
happens to be an Ace of Spades, and from what you told me any ordering of the
remaining 51 card remains possible.''

``Excellent,'' my friend replies. ``While I make some tea, please go through
the decks yourself to verify your answers.''

I diligently leaf through the two decks.  Indeed, the first deck displays the
standard ordering, while the second reveals, besides the top ace, a haphazard
arrangement---but I note that the King of Hearts still immediately follows the
Queen of Hearts.

``Was my description correct?'' my colleague asks.  ``Yes,'' I reply,
``Everything that I saw agrees with it.'' ``Then,'' he asks, ``can you tell me
their entropies again?'' ``As far as I'm concerned,'' I reply, ``the entropy is
1 for the first deck and \dots\ 1 for the second deck as well!''  ``Wait--- you
changed your mind about the second deck?'' ``Of course, my \emph{mind} has
literally changed---I definitely \emph{saw} with my eyes---and even made a
point of memorizing---what the specific ordering of the second deck was.  Of
course no other ordering would match it!''

\smallskip

We meet again a month later. He smiles and asks, ``What did you say the entropy
of the second deck was?''  ``It was 1,'' I answer, ``but as far as I am
concerned, \emph{today} its entropy is 50 factorial---or $1\by50\by1\by49!\,$''
``And why so?''  ``In a month, I forgot all about it---except that it started
with an Ace of Spades and that the King of Hearts followed his Queen.''

\xtra{That leaves just 1 possibility for the first card; only 50 possible
positions for the Queen out of the remaining 51, to make room for the King
right after her; only 1 possible position for the King, that is, right after
the Queen.  As for the 49 cards that are left, they may come in any order.}

\noindent ``You mean your entropy started with 51 factorial, then went down to
1, and now is up to 50 factorial---and all that while the deck itself was not
changing a bit?'' ``Exactly---``my'' entropy, as you say. That is, the best
description of the deck I could give myself, moment by moment, as I went
through this comedy.
 \smallskip
 \hrule

 \bigskip

The morale of this story is that the entropy ``of a system''---that is, of a
\emph{description of it}---maintained by me or any other agency---may evolve
not only when the system itself evolves (say, somebody cut the deck) but also
while the system itself \emph{remains unchanged}. It may \emph{increase}, if I
lose or forget information about it. Even if I'm very careful, it may increase
if I know that the system itself is \emph{nondeterministic} (see
\sect{nondeterministic}). It may \emph{decrease}, if I receive any information
about the system that allows me to arrive at a more detailed picture of its
makeup, or if I directly ``peek'' into it (see \sect{ehren} and
\cite{Leff90maxdemon}). As we shall see, there are also systems whose entropy
will honestly \emph{decrease} ``all by itself,'' even if the system is
perfectly isolated and no one peeks!  The second law of thermodynamics is valid
for any \emph{honest} kind of description of certain \emph{distinguished} kinds
of system (see \sect{2nd}, and especially table \eq{budget}, ``invertible''
column); as a special case, it \emph{strictly} applies to physics---as claimed
by Clausius.

\bigskip

To help turn the above intuitive considerations and assurances into something
more definite---stuff that you can trust because you \emph{created it
yourself}---in the next section we shall introduce \emph{dynamical systems},
that is, systems whose state evolves in time by well-specified internal laws.

Using dynamical systems as a tool we shall fasten the patient to the operating
table, as it were. By excising intervening layers of fat and connective tissue,
we'll lay bare with surgical precision the \emph{contractual essence} of the
second law of thermodynamics (\sect{2nd}).

 \Sect[track]{Tracking entropy}

\myquote[Benjamin Disraeli 1868]{When I want to read a book, I write one!}

\myquote[Robert Baden--Powell, the founder of Boy Scouts, who morphed in this
way, for pedagogical purposes, the previous quote]{The best way to learn a
subject is to write a book about it.}

\myquote[ttributed to Albert Einstein by John Wheeler]{If I can't picture it,
I can't understand it.}

\myquote[Richard Feynman 1998, scrawled on a blackboard shortly before his
death]{What I cannot create, I do not understand.}

\myquote[Craig Venter 2010, inscribed as a trademark on the first synthetic
working genome---in the belief he was literally quoting Feynman]{What I cannot
build, I cannot understand.}

\xtra{The above five quotations are a partial list of responses I got when I
asked my brain ``You know, that quote about how `one needs to touch it with
hand to understand it'.'' That provides a lower bound of 5 to the entropy (in
the present sense) of my clue.

By the way, I realize that the best way to learn about entropy is to write a
paper about it.}

\Subsect[dynamic]{Dynamical systems}

A \emph{dynamical system} is an abstraction of a mechanical system that can be
in any one of a set of internal \emph{states}, and is governed by an internal
\emph{rule} given once and for all---external interventions or ``miracles'' are
not allowed! Think of a music box or a planetary system (or, for that matter, a
planetarium show or a billion-year computer simulation of the solar
system---such as the one performed by Gerry Sussman and Jack Wisdom's ``digital
orrery''\cite{Sussman88}).

If the rule makes the system ``hop'' directly from one state to the next,
without going through a continuum of intermediate states (contrast a digital to
an analog watch), the system is \emph{discrete}. We shall set no upper limit to
the number of states that a discrete system may have available.  Moreover,
though not essential for our arguments, it will help intuition and simplify the
presentation to assume that the actual number of states, even though as large
as desired, be \emph{finite}.

A system is \emph{deterministic} if its rule associates to each state a single
\emph{successor} state.\Foot
 {That is, if the rule is a \emph{function} from the state set to iself.}
 For any chosen inital state, iteration of the rule produces, one after the
other, a sequence of states called the \emph{trajectory} from that initial
state.  In this sense, the rules embodies the \emph{dynamics}---or the
``behavior---of the system for all possible initial conditions.

\medskip

For the moment we shall restrict our attention to dynamical systems qualified
by the above four properties, namely, (i)~internal rules fixed once and for
all, (ii)~time-discrete, (iii)~finite-state, and (iv)~~deterministic.

 Thus, for us a \emph{system} will consist simply of a finite set of elements
called \emph{states} and a \emph{law} or \emph{rule} that to each state
assigns, as a \emph{next state} or \emph{successor}, an element from the same
set.  For example, the state set $S$ may consist of just the four states
$a,b,c,d$ (\ie $S=\{a,b,c,d\}$); and the rule $R$, which must assign to each
element of $S$ its successor, may be an arbitrary lookup table such as that on
the left of \eq{rule}, or its graphic equivalent---a \emph{state-transition
diagram}---on the right.
 \Eq[rule]{
 \renewcommand*{\arraystretch}{.5}
 \begin{array}{c|c}
 \multicolumn{2}{c}{R}\\
 \multicolumn{2}{c}{\longrightarrow}\\
 a & b\\ b & c\\ c & d \vrule width0pt height6pt\\ d & b
 \end{array}\quad\quad\quad
 \vcenter{
  \hbox{\begin{tikzpicture}
  [scale=.5,auto=left,every node/.style={circle}]
  \node (a) at (0,2) {$a$};
  \node (b) at (2,2) {$b$};
  \node (c) at (2,0) {$c$};
  \node (d) at (0,0) {$d$};
  \foreach \from/\to in {a/b,b/c,c/d,d/b}
    \draw[->,>=stealth,] (\from) -- (\to);
 \end{tikzpicture}}}
 }
 This state-transition diagram is a directed graph where \emph{nodes} denote
states and whose outgoing \emph{arcs} point to a node's successor. As we've
seen, in a deterministic system each node has a \emph{single} successor;
however, a node may well have more than one predecessor (like $b$, which has
two), or even none (like $a$). A node may be its own successor, in which case
the arc coming out of it loops back to it.

\medskip

Since in ordinary scientific speech ``state of a system'' may often mean
(especially in physics) a description of its makeup vague enough to leave room
for several individual states, as in the ``hit-and-run car'' example of
\sect{entropy001}, when necessary for clarity we may refer to the individual
elements of the state set as \emph{ur-states} (``fundamental'' or ``atomic''
states).

\xtra{Note that what should be considered an ur-state is not a given, ultimate,
intrinsic feature of a system. It is, at bottom, a \emph{modeling
decision}---what in logic is called a \emph{premise} (\sect{prob})---made for
the purpose of evaluating its consequences. Thus, the entropy of a deck of
cards discussed in the example of \sect{entropy001} is based on the number of
permutations of an ideal, indestructible deck consisisting of discrete symbols
like `{\bf K}$\spade$' and `{\bf Q}$\club$', regardless of, say a card's
orientation or smell. After a zillion shuffles this ideal deck will be just as
new, with its unblemished 52! ur-states, while a real deck will have been
reduced to a pile of dust by continual wear and tear: billions of tiny specks
susceptible to astronomically more permutations---and not even these are
ultimate states of matter!}

In the rest of this section, where ur-states are represented by nodes
in a graph, we may call them simply \emph{nodes}.

\Subsect[conserved]{Day 1: Look! It's conserved!}

Let's imagine that you keep a dynamical system well-guarded in your room. I'm
in a \emph{separate} room, and we comunicate only by intercom.

 \begin{dialog}
 \cue{You}{I have here a discrete dynamical system consisting of ten states
labeled 0 through 9.  I am going to select \emph{one} of these as an initial
node, and then I'll make the system take a few steps---that is, successively
move from one node to the next according to the system's rule. Your task will
be to identify the node in which the system is left at the end of this. Is that
clear?}
 \cue{Me}{Yes! But aren't you going to tell me what the actual system is---what
rules it obeys?}
 \cue{You}{Oops! I'd forgotten. Let me fax you the entire transition
diagram---it has nodes to represent states and arcs leading from each node to
the next. \dir{faxes diagram of \fig{eq_10}a} Got it? OK!  Now I've chosen an
initial node and put a check mark next to it. Tell me what node it is.}
 \end{dialog}

\Fig[eq_10]{\small
(a)\includegraphics[scale=.48,viewport=220 143 370 355,clip=true]{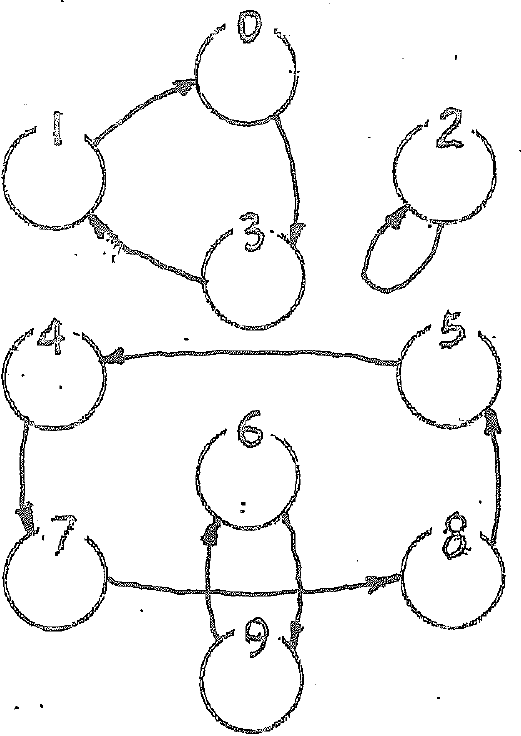}\hfil
(b)\includegraphics[scale=.48,viewport=220 150 375 362,clip=true]{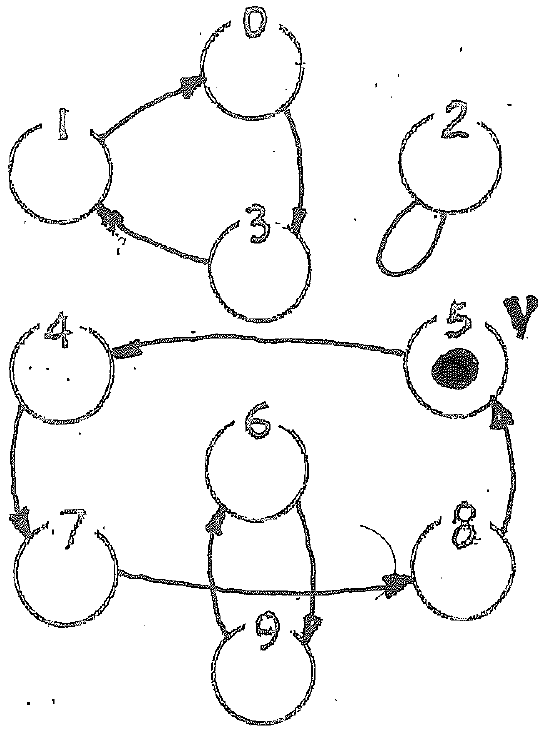}$t{=}0$
}{(a) A dynamical system, in (b) an initial state.}

 \begin{dialog}
 \cue{Me}{How would I know? It could be any of the ten.}
 \cue{You}{That's right---I haven't told you which one I've chosen. Even though
\emph{I} know what node I have chosen for the initial node, and so the entropy
of \emph{my} description (\emph{how many} nodes the system could possibly be in
after I'd made my choice) equals 1, nonetheless the entropy of \emph{your}
description---your ``knowledge state'' given what you've been told so far (and
you can't see my check mark)---is still ``any of them ten.''\\
 \null\qquad Now I tell you what initial node I've chosen: it's 5 (your entropy
as to the initial state must surely be 1 now). My command ``Step!'' will mean I
advance the system's node by one step. Are you ready? Then let's start!}
 \cue{Me}{Just a moment! \dir{puts a checkmark next to initial node 5, for the
record, and places a movable token on the node itself, as in \fig{eq_10}b} OK, shoot!}
 \cue{You}{Step!}
 \cue{Me}{\dir{advances the token by one step; the token is now on node 4}
OK!}
 \cue{You}{Step! Step! Got it?}
 \cue{Me}{Fine! \dir{advances the token two more times}}
 \cue{You}{Take four steps this time! I'm done. What's my final state?}
 \cue{Me}{\dir{takes four more steps on the graph and draws a dotted line on
the diagram to trace the token's entire trajectory, as in \fig{eq17}; the token
has landed on node 8} You must be now on node 8!}
 \cue{You}{Exactly! And I see that your entropy is still 1---you have no doubts
about which state my system is in \emph{now}, even though you don't see it!}
 \end{dialog}

\Fig[eq17]{\small
\includegraphics[scale=.48,viewport=220 146 375 355,clip=true]{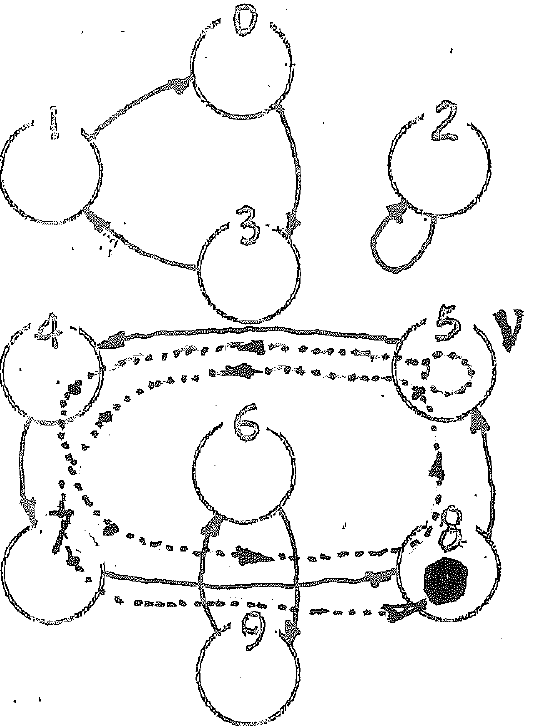}$t=7$
}{Final state (after seven steps) of the given law.}

 \medskip\centerline{*****}\medskip

\noindent A little later---same setting.

\smallskip

 \begin{dialog}
 \cue{You}{This time I'm not going to tell you exactly which node I've chosen
as the initial state---just that it is one of these: $\{1,4,7,9\}$. Ready?}
 \cue{Me}{Just a moment! \dir{marks nodes 1, 4, 7, 9 on his diagram, for the
record, and places a token on each, as in \fig{eq407}a} Then my initial entropy
is 4. I guess I'll have to treat the four initial nodes as separate cases, and
at every step advance the four tokens ``in parallel.''  Why not? it's pretty
much like if I were managing four game boards at once. Go ahead, shoot!}
 \cue{You}{Step! Step!}
 \cue{Me}{\dir{carefully moves each of the four tokens two steps forward, all
on the same diagram sheet} OK---but don't go too fast!}
 \cue{You}{Now just do nine more steps---take your time! That will be all.}
 \cue{Me}{\dir{is done after a little while, and at the end also records the
\emph{final} arrangement of tokens on his worksheet---as in \fig{eq407}b} The
final state may be any of these nodes: $\{3,4,5,6\}$}
 \end{dialog}

\Fig[eq407]{\small
(a)\includegraphics[scale=.48,viewport=220 142 375 355,clip=true]{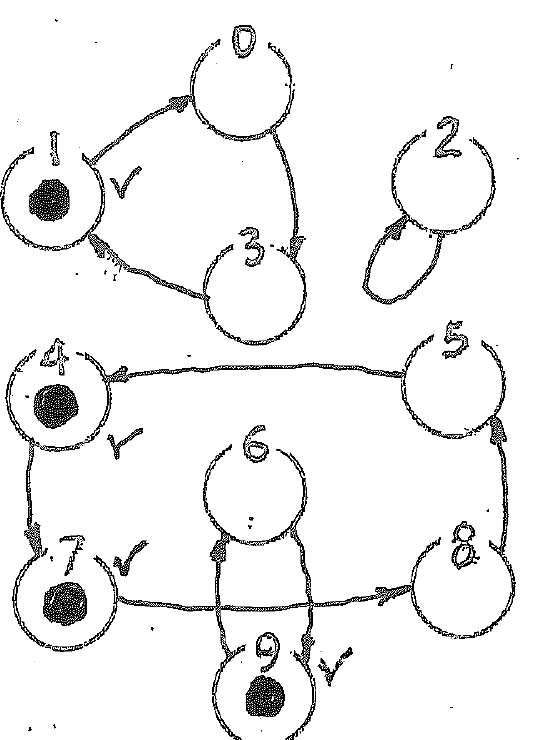}$t{=}0$\hfil
(b)\includegraphics[scale=.48,viewport=216 145 375 355,clip=true]{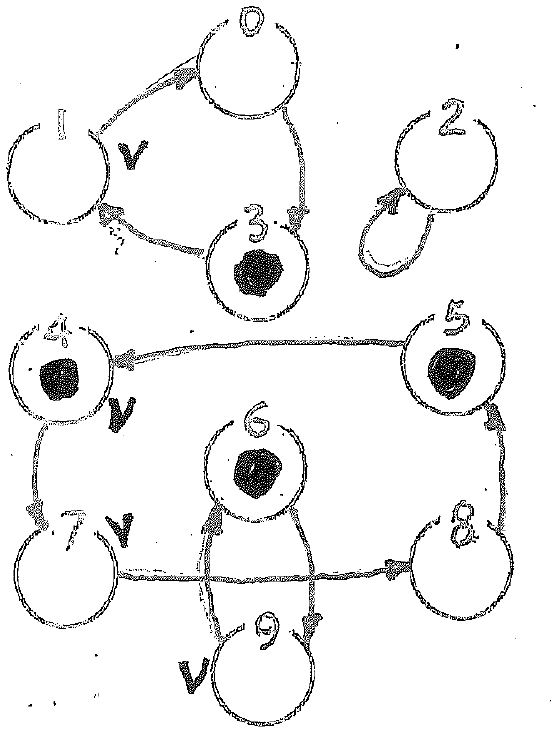}$t{=}11$
}{Initial ($t{=}0$) and final ($t{=}11$) state of the same system, but run with four tokens.}

\medskip

You now invite me to your room to compare the outcome of my ``duplicate play''
with the final state of the ``master play'' in your room. The results agree, in
the sense that (a) the final resting place of your (single) token matches one
of the (four) places that are covered by a token in \emph{my} board, and (b)
there are \emph{no more} tokens on my board than would be necessary to
guarantee that result for \emph{any} of the initial nodes you'd stated as
possible! 

Also, I remark that in this case as well my entropy at the end of my
``remote-control'' performance is still equal to that of the initial state.
(Of course it then collapsed from 4 to 1 as soon as I saw \emph{your} actual
final node.)

 \medskip\centerline{*****}\medskip

\Fig[eq99]{\small
 \includegraphics[scale=.48,viewport=220 142 375 355,clip=true]{fig/eq_.pdf}\hfil
 \includegraphics[scale=.48,viewport=216 145 375 355,clip=true]{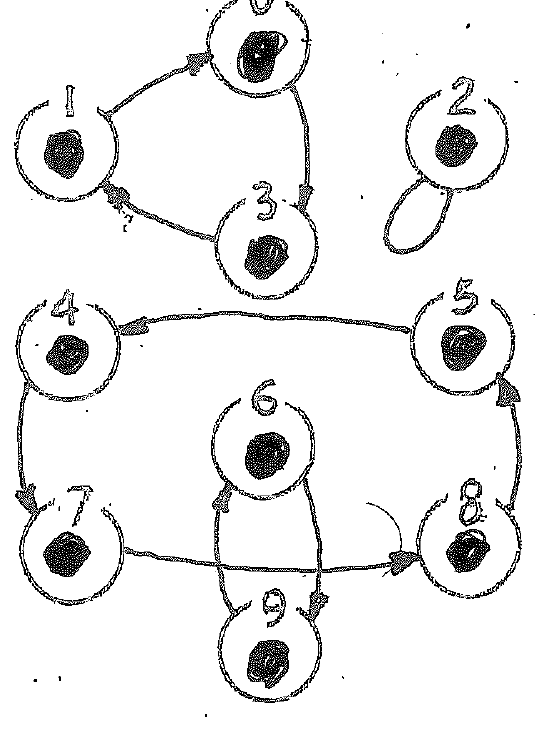}$t{=}0$
 }{Blank system template and, on its right, an initial state consisting of a
fully-filled template.}

 \noindent Still later---same setting.

\smallskip

 \begin{dialog}
 \cue{You}{Look, I'm tired of choosing initial states. Let's say that I may
start from \emph{any} node. Ready?}
 \cue{Me}{\dir{dutifully marks all ten nodes as possible candidates for the
initial state, and covers each with a token, for a total of ten tokens, as in
\fig{eq99}} I'm ready!}
 \cue{You}{Then I'd like you to take \emph{eleven} steps of it!}
 \cue{Me}{I will---but with all these tokens it's going to take a while!
\dir{moves each of the ten tokens one step forward} One! \dir{a second step} Two!
\dots\ \dir{notices that in spite of all that bustle nothing has changed as a
result; looks again at the state diagram; then, in a flash of understanding}
Done!}
 \end{dialog}

\smallskip\noindent What's happened? By inspecting the state diagram of this
particular dynamical system, I noticed that it consists of just \emph{loops},
or closed trajectories. Not only does each node have exactly one successor, but
also exactly \emph{one predecessor}! Thus it never happens that two arcs flow
into the same node, or that a node has no inlets at all. The dynamics (\ie the
system's ``next state!''  law) is a \emph{one-to-one correspondence}; in other
words, it is \emph{invertible}. That is, you could go backwards and
unambiguously retrace your steps by following the arcs \emph{backwards}. When
stepping forward on this diagram, a ``platoon'' of $n$ tokens occupying $n$
distinct nodes will invariably land again on $n$ \emph{distinct} nodes---there
are no ``merges'' of trajectories, no piling-ups of tokens onto the same node.
The number of occupied nodes---the entropy of the platoon's
``footprint''---does not change as the platoon advances---no matter whether by
one step or one billion, or whether you start with a single token (an
ur-state), a small patrol of tokens, or a whole platoon! If we take ``entropy''
to mean ``amount of information,'' an invertible rule is (ach!)
\emph{informationlossless}---whether the system's description is \emph{precise}
(it pinpoints a single ur-state) or \emph{vague} (leaves one uncertain as to
which of a set of many ur-states).

In the present case, since \emph{all} nodes were occupied by a token at the
beginning of the evolution, as in \fig{eq99}, they will remain so after each
step. Since my task is to record, on the basis of the initial description and
the number of steps taken afterwards, the nodes possibly occupied \emph{now},
and \emph{all} nodes are ``possibly occupied'' at the start, the conclusion is
that \emph{all nodes} are possibly occupied at \emph{any time}. In sum, even as
the tokens advance, their occupation pattern does not change at all with
time---and that's why I can announce the final distribution instantly!

By the way, if in spite of the movement of tokens the overall token pattern is
found to be the same at the end of each step, then this pattern is called an
\emph{equilibrium state}---another way of saying time-invariant or
time-independent.

\Subsect[poormans]{A ``poor man's'' MaxEnt principle?}

Note that, in any of these duplicate plays, I strive to cover with tokens
\emph{every} node for which I had some reason to think that they (but no
others) \emph{might} be occupied in your master play---as remarked at the end
of Day 1's second dialogue (\fig{eq407}). What is the rationale for that?

\medskip

In the Jaynes entry of \sect{entropy000}, we saw that the MaxEnt principle is
but the implementation, in a statistical context, of a witness's categorical
imperative of \emph{honesty} (``the whole [ascertainable] truth and nothing but
the truth'').

But in the present lean-entropy context, where no \emph{statistics} has yet
been introduced, the only tool that forecasters have at their disposal to
implement that categorical imperative, in organizing and updating their
evidence, is the blunter ``all-or-nothing'' of \emph{logic}.  Here, the ``whole
truth and nothing but the truth'' clause is simply expressed by a \emph{sum of
minterms}, ie, by the logic {\sc or}ing of {\sc and} clauses; one may call
this approach \emph{MaxMin}.  The result is a \emph{subset} of the set of all
conceivable ur-states (aka `outcomes'), or, equivalently, the
\emph{characteristic function} of that subset, which assigns to each outcome
the stark choice between 1 and 0 (for `currently possible' and `currently not
possible'). Contrast this with a probability distribution, which may assigns to
each outcome a \emph{weight} anywhere between 0 and 1: for the same honest
intention, if you are given finer tools you \emph{may} get sharper results.

\medskip

What I want to stress is that the MaxEnt principle is not sacred: the ultimate
criterion is that of \emph{honesty}, of which MaxEnt is a derivative---the best
implementation made possible by a certain kind of evidence and certain updating
(or ``evidence propagation'') tools. Even when the introduction of statistics
will enable us to use MaxEnt instead of the present MaxMin, we may still not be
doing the theoretical \emph{best} in our inference task. That can only come by
tracking the progress of ur-states by literally applying to the ur-state set
the ``ur-law'' that maps ur-states to ur-states---no matter whether the state
of a system is described as a single ur-state; by a subset of ur-states or a
probability distribution over the set of ur-states; or other reasonable
means. This concern is captured by the concept of \emph{internal entropy},
defined in the next section.

\Subsect[micro]{Internal entropy}

The morale of Day 1's dialogs is the theorem\Foot[proof]
 {The dialog itself is what one might call a ``proof by example without loss of
generality.''}
  that
 \Law[internal]{the internal entropy\\ of an invertible dynamical system\\
 is constant:\\it never decreases or increases.
 }

\noindent What's this ``internal entropy'' stuff? Just a physicist's shorthand
for what we've been practicing in the last three pages---that is, entropy as
tracked by a \emph{certain kind} of discipline. Namely, given a description (a
collection of ur-states) of a system's initial state, \emph{internal entropy}
is the entropy (the ur-state count) of this descriptions as it evolves strictly
on the basis of a complete characterization of the subsequent
transformations---how many steps and of which law.

At present, a state is described as a collection of ur-states, and, if the law
is a many-to-one mapping (as in the dialog of \sect{decrease}), when two or
more trajectories \emph{merge} we shall apply the ``MaxMin'' honesty criterion
(see \sect{poormans}): the new occupation contents of a node is the {\sc or}
(or \emph{logic sum}) of the contents of the predecessor nodes.  But the above
definition of \emph{internal entropy} applies just as well to
probability-weighted ur-states, in which case a merging of trajectories gives
as a resulting weight the \emph{arithmetic sum} of the predecessors' weights.

\medskip

What I termed ``internal entropy'' is often loosely called ``microscopic'' or
``fine-grained'' entropy. The latter terms are a historical legacy---they have
nothing to do with \emph{very small systems} per se. What can there be of
\emph{microscopic} about, say, a deck of cards? On the other hand, as this deck
evolves according to the dynamics of, say, a \emph{zero-person} solitary game,
in which the player has no choice about how to move since all the moves are
\emph{forced} by the rule---and we are thus dealing with a \emph{dynamical
system}, its entropy (as a measure of my \emph{best achievable} current
description) evolves as well. This \emph{ideal} process of accounting for
entropy evolution is \emph{uniquely} determined by the ``internals'' of the
system (\ie the objective rules of the solitary game), and not by any demands
or happenings ``external'' to the system, like my (possibly non-ideal)
accounting.

\xtra{``If a tree falls in a forest and I was listening to my iPod full-blast,
does the tree make a sound?'' I may swear I heard no sound. But, in falling,
the tree left marks that \emph{in principle} any one can later investigate to
determine what happened in the forest. If sufficient efforts are made,
different investigators will come to the same conclusions, thus reconstructing
an objective ``internal'' happening independent of my subjective, possibly
faulty, ``external'' testimony.}

In brief, internal entropy is that whose changes are wholly determined by the
ur-state transitions prescribed by a system's rule.  Note that
from the present viewpoint, in agreement with Clausius's intuition,
internal-entropy \emph{changes} are to be associated with
\emph{transformations}.  External-entropy changes (discussed in \sect{2nd}\,ff.)
are also associated with tranformations, but of a different kind.

\medskip

The key feature of internal entropy is the assumption of \emph{complete
knowledge} of the \emph{rules}---such as the transition diagram of the above
dialogs---\emph{as they apply to the system's ur-states}. It doesn't matter
instead whether a \emph{description} of the state of the system is fine or
coarse---whether it pinpoints a ur-state (as in the first dialog), leaves open
a number of possibilities (a ``cloud'' of ur-states, as in the second dialog),
or is totally vacuous (as in the third).

Once we know the rule, we can in principle precisely apply it to \emph{any}
individual ur-state. The collective \emph{makeup} of an ur-state cloud---which
represents the ambiguity of our state of knowledge---is \emph{irrelevant}; and
so is (in principle) the amount of labor needed to apply the law to
\emph{every} ur-state.

\smallskip

\xtra{Note that the token-pushing labor can be distributed among many
workers and performed by them concurrently, independent of one another, since
tokens do not interact. As a matter of fact, in certain situations the image of
a \emph{cloud} of tokens crowded over a single state-diagram may not have
adequate expressive power (see \sect{increase}). To play it safe, one should
always imagine using as many instances of the game board---as many identical
copies of the state diagram---as there are tokens, and have each token run
\emph{all by itself} on its own board (see \sect{many}).}

\Subsect[decrease]{Day 2: It goes down!}

 \Fig[dn_0]{\small
(a)\includegraphics[scale=.48,viewport=218 150 370 364,clip=true]{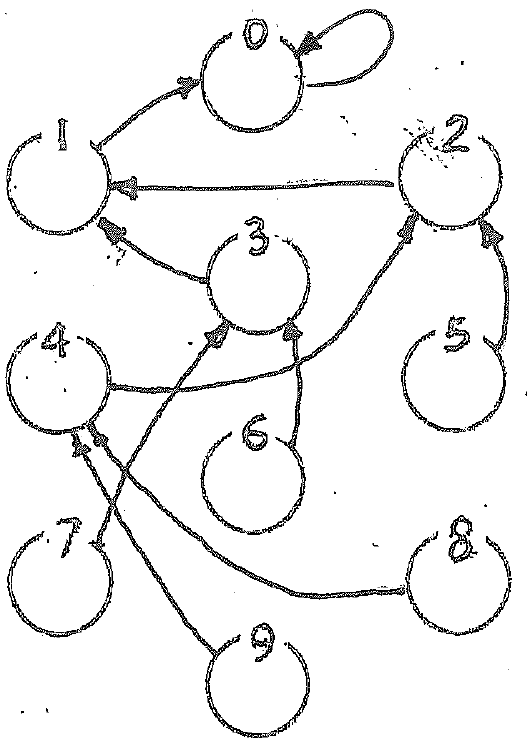}\hfil
(b)\includegraphics[scale=.48,viewport=215 150 370 364,clip=true]{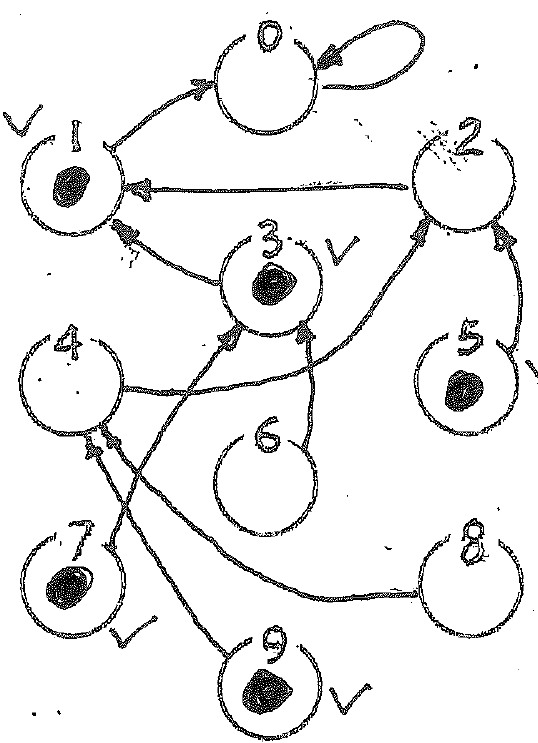}$t=0$
 }{(a) Another dynamical system, with initial state (b).}

\begin{dialog}
 \cue{You}{I just faxed you a new dynamical system---same ten states as the
old, but the arcs are different \dir{see \fig{dn_0}a}.  Wanna play a game?}
 \cue{Me}{Sure! What's up?}
 \cue{You}{I'm asking you to tell \emph{me}! I suspect there may be something
different. Let me describe the initial state as ``an odd number.''}
 \cue{Me}{You mean, any of $\{1,3,5,7,9\}$?  I'll have to use five
tokens. \dir{see \fig{dn_0}b} Let me make your system go through a few steps and
I'll tell you what happens.}
 \cue{You}{Go ahead!}
 \cue{Me}{Well, on the first step the token moved forward like so: $9\to4$,
$7\to3$, $5\to2$, $3\to1$, and $1\to0$. The occupancy is now $\{0,1,2,3,4\}$;
the entropy level has remained at 5. Incidentally, the rule seems to be
``divide by two and throw away the remainder.''}
 \cue{You}{Take one more step!}
 \end{dialog}

 \Fig[dn12]{\small
(a)\includegraphics[scale=.48,viewport=216 148 370 360,clip=true]{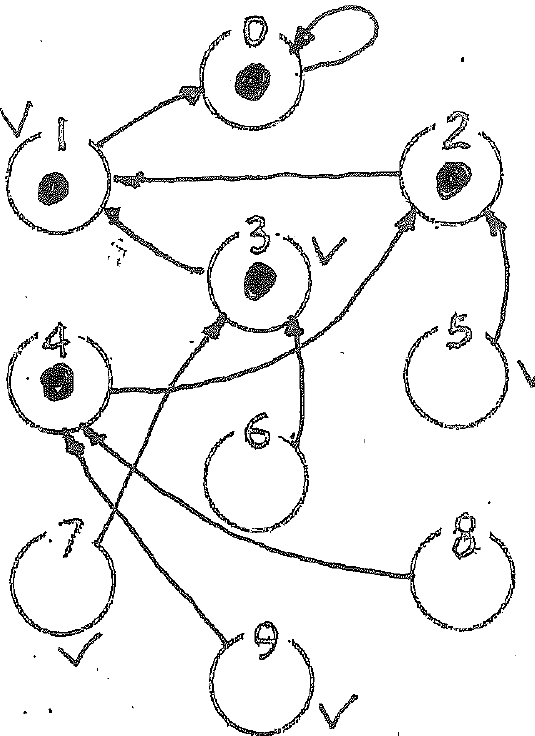}\hfil
(b)\includegraphics[scale=.48,viewport=218 150 370 364,clip=true]{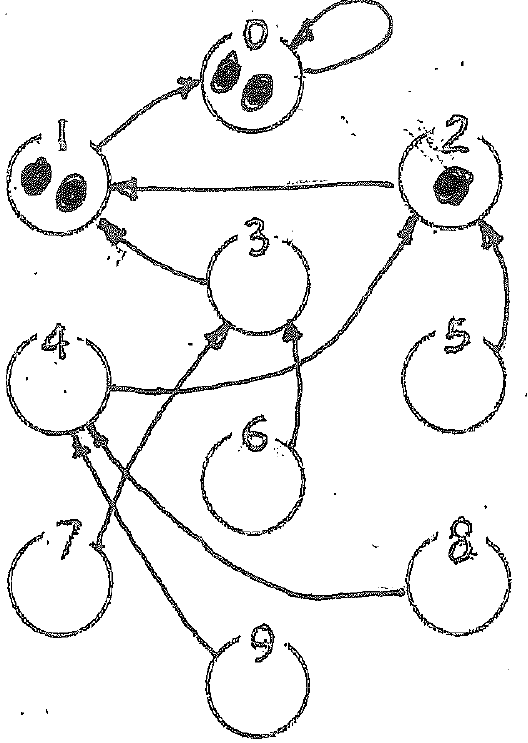}
 }{Same at times 1 and 2. In (a), both tokens 2 and 3 prepare to land on
node 1, and both 1 and 0 on node 0!}

 \begin{dialog}
 \cue{Me}{One moment! While the token on node 4 \dir{see \fig{dn12}a} wants to
go on node 2, both 3 and 2 want to go to 1, and both 1 and 0 to 0. What shall I
do with this paired tokens?---allow a double occupancy and place them both on
their destination node?}
 \cue{You}{The way we've been using tokens so far is to tell, ``Given the
system's dynamics and the description of its initial state, there is a
possibility that the system will be on this node.'' With this interpretation at
least, it doesn't make a difference if you move both tokens to the node they
want to occupy, or put only one token there and take the other off the board.}
 \cue{Me}{I'll do the former---in all these ``overbooking'' cases I'll ``put
two or more passengers in the same seat!'' \dir{see \fig{dn12}b} Now there are
tokens only at $\{2,1,0\}$; entropy has dropped from 5 to 3.}
 \cue{You}{Go on. Step! Step! Step!}
 \end{dialog}

 \Fig[dn34]{\small
(a)\includegraphics[scale=.48,viewport=218 148 370 360,clip=true]{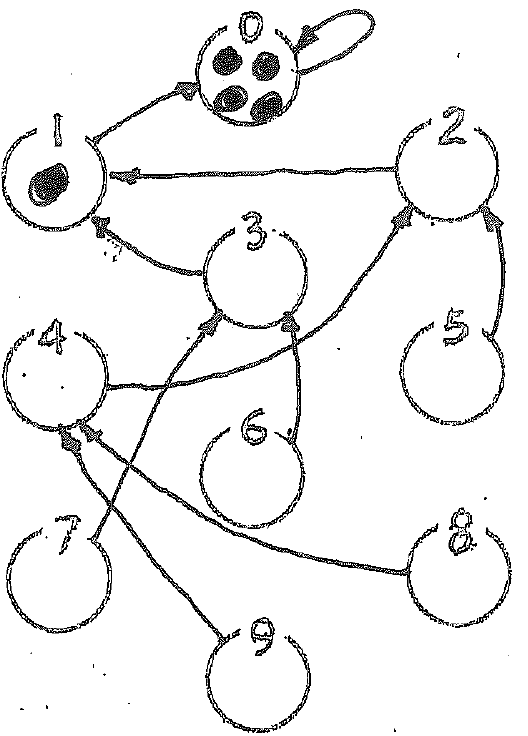}\hfil
(b)\includegraphics[scale=.48,viewport=216 148 370 360,clip=true]{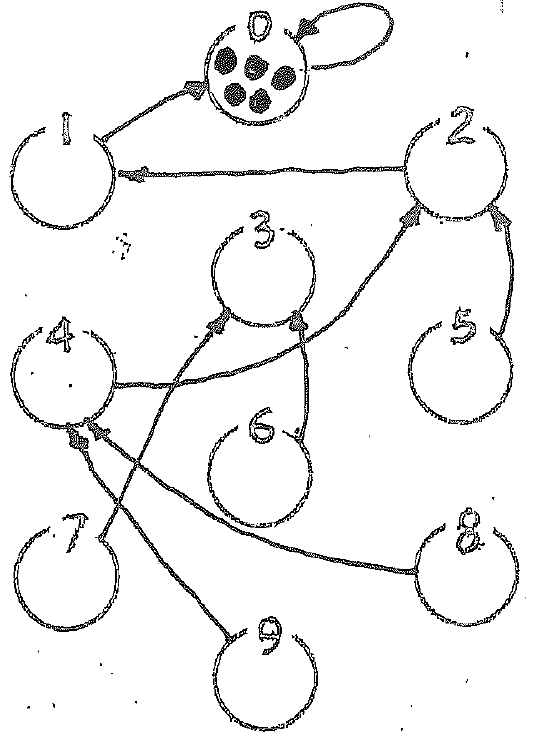}
 }{Same after steps 3 and 4. By now, all tokens have flown into node 0,}

 \begin{dialog}
 \cue{Me}{Occupancy dwindled down first to $\{1,0\}$, then to just
$\{0\}$---and that's where things will remain from now on. In fact, once a
token has reached node 0, in one step it will cycle back to 0---and so on
forever! It will never go off the board!  Entropy went down to 2, then to 1,
and now it is stuck there.}
 \end{dialog}

 \Fig[partialorder]
  {\includegraphics[scale=.5,viewport=245 140 360 350,clip=true]{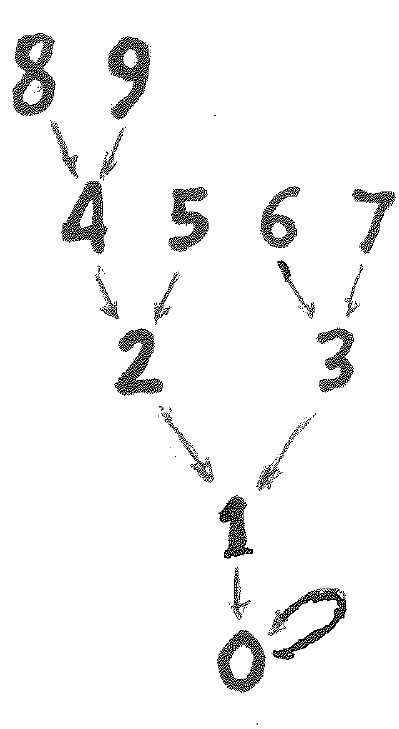}
 \small\renewcommand*{\arraystretch}{.7}
 \let\bul\bullet
 \begin{tabular}[b]{c|cccccc}
 \multicolumn{1}{c}{node} & \multicolumn{6}{c}{state occupation} \\
 9 & $\bul$ \\
 8 \\
 7 & $\bul$\\
 6 \\
 5 & $\bul$\\
 4 &            & $\bul$\\
 3 & $\bul$     & $\bul$\\
 2 &            & $\bul$ &$\bul$\\
 1 & $\bul$     & $\bul$ &$\bul$ &$\bul$\\
 0 &            & $\bul$ &$\bul$ &$\bul$ &$\bul$ &$\bul$\\\hline
 \multicolumn{1}{c}{entropy}
        & 5 & 5 & 3 & 2 & 1 & 1 \vrule width0pt height9pt\\\hline       
 \multicolumn{1}{c}{$t\longrightarrow$}
        & 0 & 1 & 2 & 3 & 4 & 5 \vrule width0pt height8pt
 \end{tabular}
 }
 {(left) Rearranged transition diagram to highlight the
partial ordering of states. (right) Time graph of node occupation and
corresponding entropy.}

 \begin{dialog}
 \cue{You}{So there were steps when the entropy didn't change; and others
when it went down; but none when it went up?}
 \cue{Me}{That's right, the entropy of your system monotonically
\emph{decreases} as time goes on, and I can prove it!  In this diagram here
\dir{see \fig{partialorder} left} I rearranged nodes and arcs to make explicit this
inexorable march of events.}
 \cue{You}{That's strange \dots\ physicists say that internal entropy can
\emph{never} decrease: if it did, one could exploit that to turn waste heat
into useful mechanical work---to make a perpetual-motion machine!  What is
different with \emph{this} system?}
 \cue{Me}{Well, physics is a special case, since its fundamental laws---whether
in the classical or quantum-mechanical formulation---are believed to be
\emph{strictly invertible}. But the new system you gave me to play with is
\emph{not} invertible! \dir{reassuringly} Don't be embarrassed---I hear that
the noninvertible ones are actually the majority!}
 \cue{You}{I'm satisfied with your explanation. But physicists also maintain
that entropy keeps increasing all the time---only in ideal circumstances does
it remain constant. Yet in invertible systems, as we saw in Day 1, internal
entropy is strictly \emph{constant}, and in noninvertible ones it sometimes
\emph{goes down}. Between invertible and noninvertible---that takes care of
\emph{all} systems, and so leaves no room for entropy ever to go \emph{up}.  It
seems that the difficulty is how to make entropy go \emph{up} rather than down!
How do physical systems manage the trick of letting entropy increase in spite of
their being governed, as you just said, by \emph{invertible} laws?}
 \end{dialog}

\Subsect[increase]{Day 3: It goes up!}

Today you walk into my room with a glum face.
 \begin{dialog}
 \cue{You}{My fax machine is broken.  But I still don't want you to peek into
the system in my room---after all you are supposed to make \emph{forecasts},
not reports. In lieu of faxing, I will chalk the transition table on a slate
and give \emph{that} to you.}
 \cue{Me}{Fine with me!}
 \cue{You}{\dir{scribbles on the slate} Here is the system's dynamics!  Can you
run seven steps of it? Start with $\{4,7,9\}$ as the initial
state. \dots\ Now---I'm sorry---I have to go see my doctor. I'll be back after
lunch.}
 \end{dialog}

\medskip
\Fig[up_0]{\small
\includegraphics[scale=.48,viewport=220 143 370 355,clip=true]{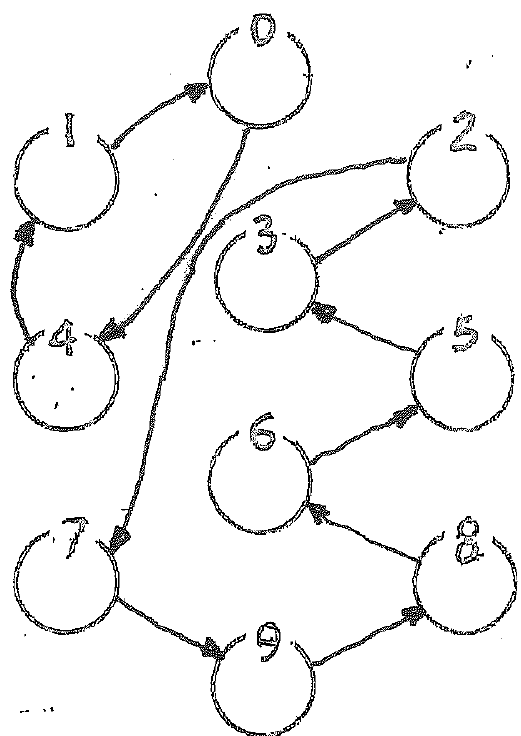}\hfil
\includegraphics[scale=.48,viewport=215 143 375 362,clip=true]{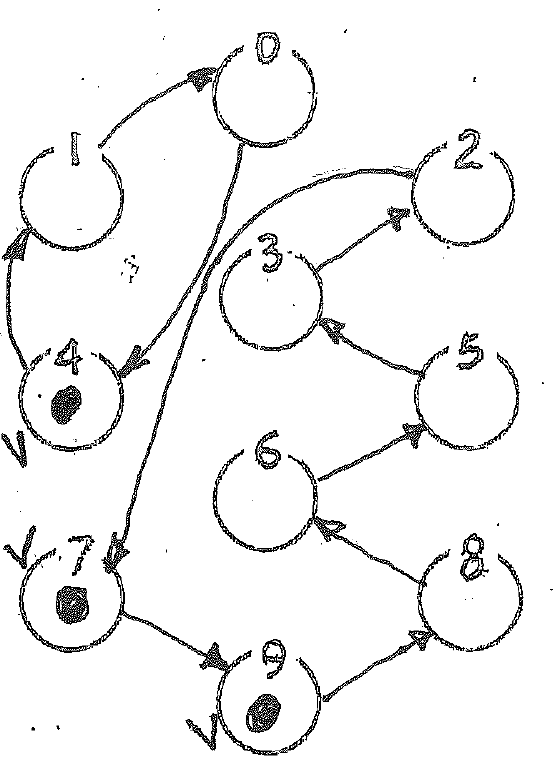}$t{=}0$
}{(a) A new dynamical system, and an initial state for it.}

I pick up the slate (\fig{up_0}). On it there is a graph with ten nodes and ten
arcs---as in the previous two days. I mark nodes $\{4,7,9\}$ as the initial
state and place one token on each. The inital entropy is clearly 3.

\Fig[up123]{\small \includegraphics[scale=.48,viewport=218 143 370
355,clip=true]{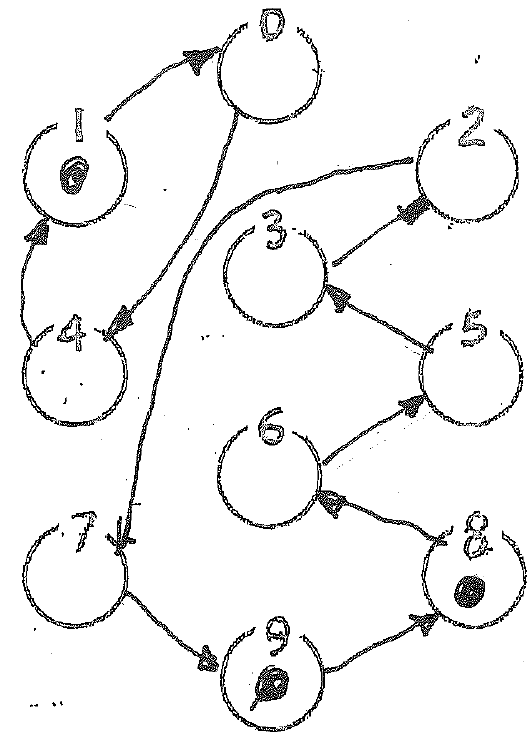}1\hfil \includegraphics[scale=.48,viewport=218 143
375 362,clip=true]{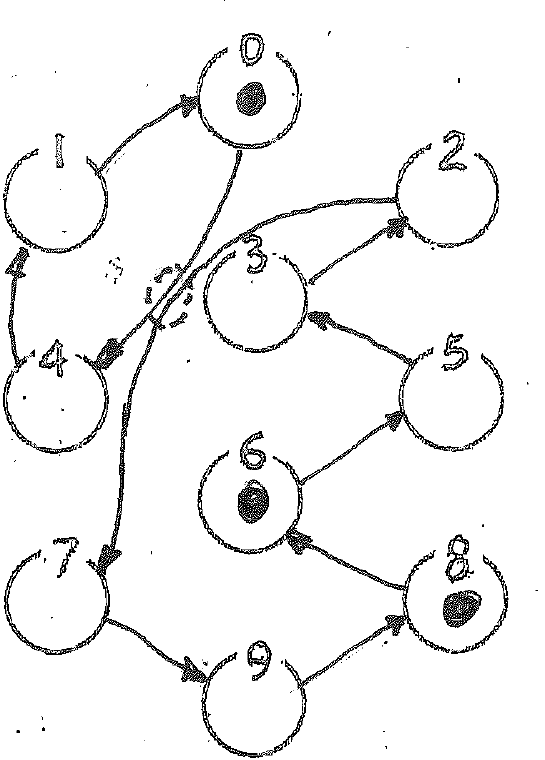}2\hfil \includegraphics[scale=.48,viewport=218
143 375 362,clip=true]{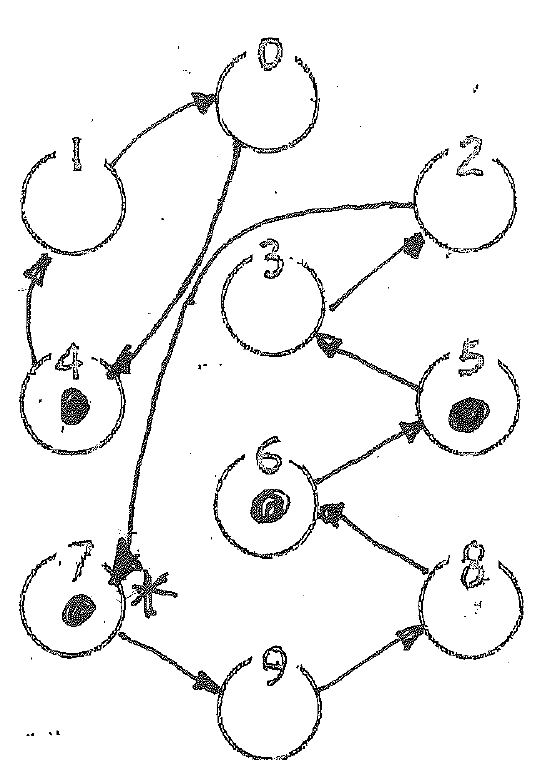}3 }{A problem in stepping between times 2
and 3: the two arcs from nodes 0 and 2 run so close to one another (where
indicated by the small dotted circle at time 2) that's hard to see whether they
run alongside one another or cross paths.}

I observe that no two arcs converge onto the same node, so the system is
invertible---as that of Day 1. I run two steps without any problems. But at the
third step I see a problem: When one of my tokens happens to be on node 0, as
at time 2 of \fig{up123}, its outgoing arc at a certain point runs so close to
that coming out of node 2 that's hard for me (I left my reading glasses at
home) to tell whether at that point the two arcs \emph{run alongside} each
other or \emph{cross}. In my case, if the two arcs run parallel my token move
should be $0\to4$; if they cross, $0\to7$.  Note that the ambiguity in this
scenario is not as to \emph{which node}, as when we gave a choice of several
nodes for the initial state, but as to \emph{which destination} accroding to
the rule---whether the transition table for node 0 says $0\to4$ or $0\to7$.

Since my task is to determine, at every step, on what nodes there can be a
token---that is, which system states are possible according to the information
I have---the best I can do at the moment is entertain both possibilities at
once for the token now leaving node 0, by placing tokens on \emph{both}
possible destinations---nodes 4 and 7.

My approach means, in substance, ``when in doubt, track the consequences
of all plausible alternative hypotheses.'' 

\xtra{When a ``game'' is played as a response to a real-life challenge, a token
may mark, say, one of the places where a terrorist might be at the
moment---remember the hit-and-run story? At the next step of the game
\emph{that} position is updated, say, according to the presumed speed of the
vehicle and the nature of the terrain---and that is done ``in parallel'' for
all alternative routes that terrorist may have taken.

\smallskip

\noindent The rationale for choosing to act that way runs much like
this. Suppose I forgot where I left my glasses and I have to instruct somebody
to go get them for me. I am to give them a list of places where they should
look. If one place which comes to my mind is not after all so likely, should I
include it? If I do, the errand will take a little longer\Foot
 {Or merely just as long, if they find my glasses \emph{before} looking at that
place.}
 If I don't include it, and the glasses actually happen to be in that unlikely
place, then my helper will come back after the \emph{longest} possible time
(having gone through \emph{all} the places in the list) and still return
\emph{empty-handed}, adding, as it were, insult to injury! I'd rather play it
safe.}

\Fig[up4567]{\small
 \includegraphics[scale=.38,viewport=218 143 370 355,clip=true]{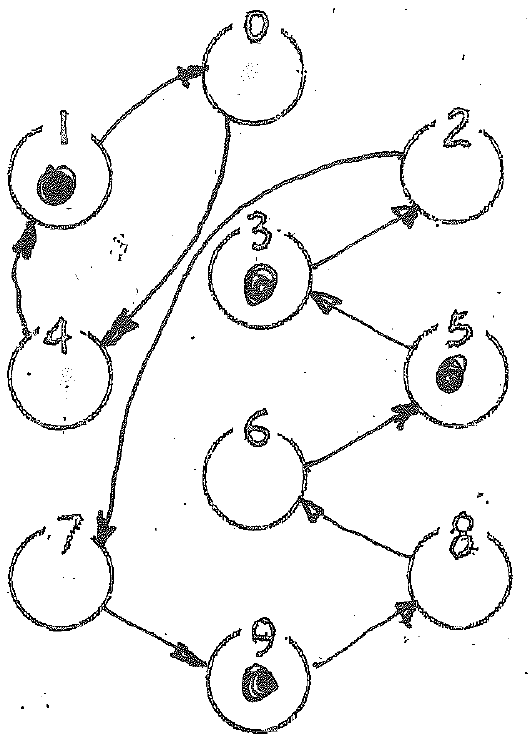}4%
 \includegraphics[scale=.38,viewport=218 143 375 362,clip=true]{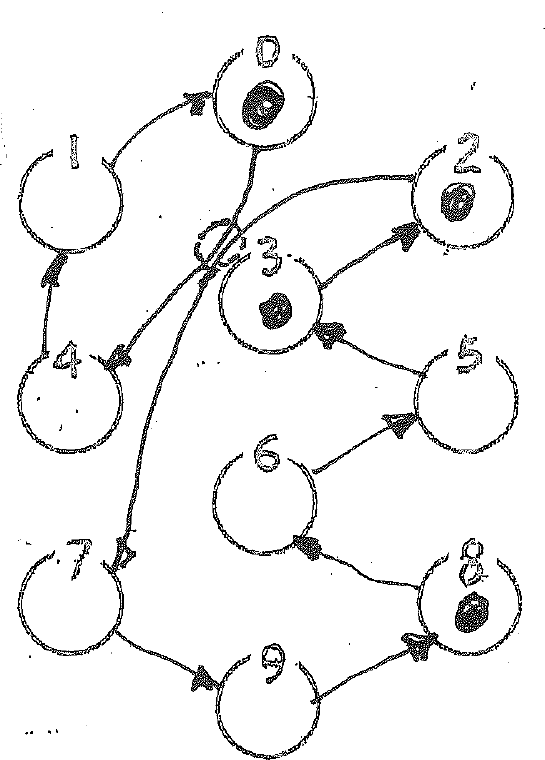}5%
 \includegraphics[scale=.38,viewport=218 143 375 362,clip=true]{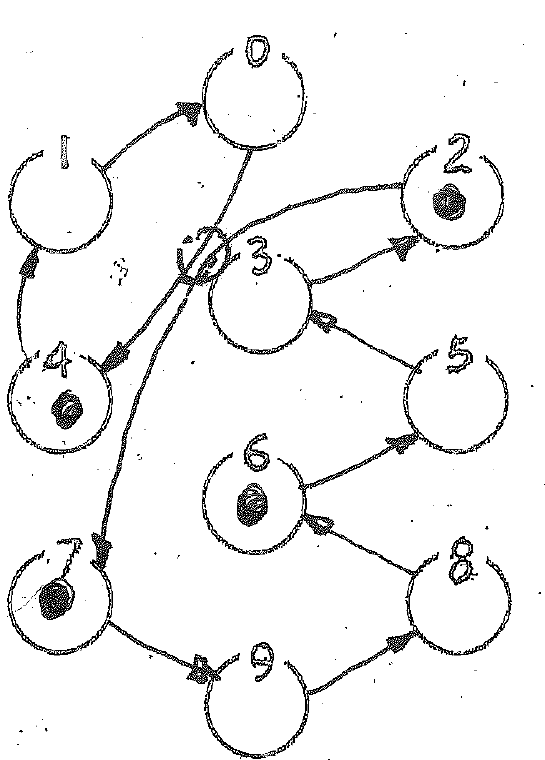}6%
 \includegraphics[scale=.38,viewport=218 143 375 362,clip=true]{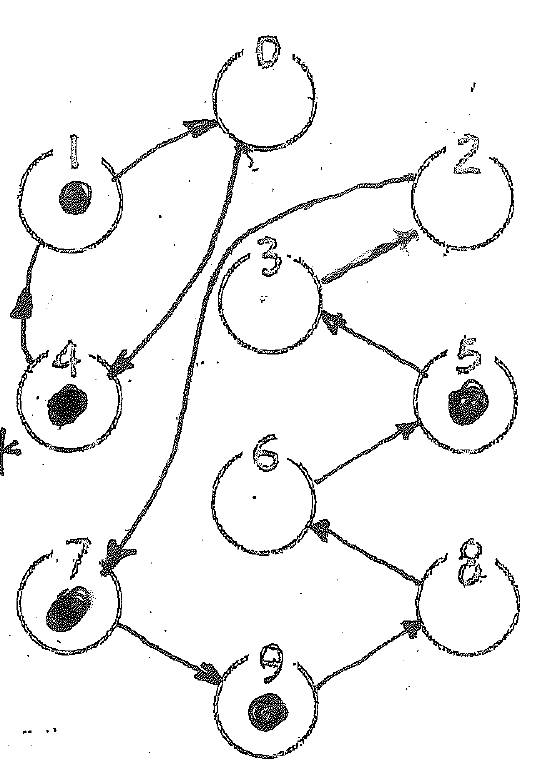}7
 }{The same problem at times 5 and 6---are the paths from nodes 0 and 2
crossing or running parallel? But at time 5 there is a token at both nodes, and
so it doesn't matter which goes where. At time 6, however, only one token is
present, and which way it goes matters.}

I go all the way through the remaining four steps (\fig{up4567}). At time 5 I
have a similar problem as at time 2 (\fig{up123}), but now \emph{both} nodes at
the head of the converging arcs have a token, so it doesn't matter whether the
two arcs cross or go along.  At time 6, only one token is present, at node 2,
and again---as at time 2---the token at node 2 is confronted with two
alternative paths and ``in doubt'' it takes \emph{both}. I have now five
tokens!

\medskip
You come back from the doctor.

 \smallskip
 \begin{dialog}
 \cue{You}{Well, how did it go? Where are your tokens now?}
 \cue{Me}{At nodes $\{1,4,5,7,9\}$; the entropy is now 5.}
 \cue{You}{How can that be? I'm sure I had given you an \emph{invertible}
system, and thus the entropy should have remained constant. Or perhaps you
believed I had given you \emph{five} alternative nodes for the initial state?}
 \cue{Me}{No, I know you only gave me three. But, owing to its poor
readability, the graph you'd given me confronted me with two alternative
routings.  Effectively, instead of a single law I had two alternative laws to
deal with.  I did my best, and kept both alternative universes  alive. My final
number of tokens represents the logic {\sc or}---the ``merge''---of tokens from
two parallel universes after seven steps of running.}
 \cue{You}{I guess you did your best---all this is very helpful. Thank you!  In
fact---please hang around---I have one more task in mind for you!}
 \end{dialog}

 \smallskip

\xtra{Note that the {\sc or} operation (the ``merge'' of token patterns) is a
\emph{hack}---a simple but not necessarily the best way to extract a token
count from the parallel evolution of a number of universes.}

\bigskip

Here we've seen one way that entropy can go up. From Day 1 to Day 3 nothing has
changed with the nature of the system's law or of its initial state, but an
accident of misplaced reading glasses prevented me from making out one bit of
the law itself---``Do these two paths cross or run through side-by-side?''
During the step where one token went through that juncture, the ``internal
law'' that governed the movement of \emph{your} tokens had not
changed---specifically, it was still \emph{invertible}---but because of some
``external perturbation'' \emph{my description} of it went through a ``forked
path.'' To represent the two alternative universes, my hack was to duplicate
the token while remaining on a single game board. As a variant of that, I could
instead have cut the token in half and let the two halves follow separate
courses. The most \emph{conservative} solution---I use that term
literally---would have been to duplicate the whole setup, board and tokens,
\emph{at the very beginning}, when I realized I didn't have
 my glasses; use the ``across'' variant of the law for one setup and the
``along'' variant for the other; and proceed with the simulation on both
boards. Later, when I found my glasses, I would have resolved the doubt and
accordingly simply discarded the wrong variant.

\xtra{You may argue, ``What a waste of resources!'' But this is exactly what is
done today by employing \emph{speculative execution} in microprocessors, where
for the sake of reducing latency a computer system performs some task that may
later turn out not to be needed. When the {\sc cpu} encounters a conditional
statement---``{\sc if} $\langle$condition$\rangle$ {\sc then do}
$\langle$this$\rangle$ {\sc else} $\langle$that$\rangle$,'' and the
value---{\sc true} or {\sc false}---of the condition is not yet known, it starts
running through both ``horns'' of the dilemma---$\langle$this$\rangle$ and
$\langle$that$\rangle$---in parallel, using extra resources provided for the
purpose. Some time later, when the condition's value is known, one horn is
lopped off and the resources already spent on it go down the drain. In this way
we save \emph{time}, since the ``right'' answer starts being computed (together
with the ``wrong'' one) before the condition is known---but at the price of
``haste makes \emph{waste}!''}

No matter what hack I use, \emph{my accounting} of the process becomes in
effect \emph{nondeterministic} (see \sect{nondeterministic}). Note I was careful
not to say ``stochastic'' or ``probabilistic''---that is, I didn't give
\emph{weights} to the two alternatives (that would be a higher-level hack,
touched on briefly in \sect{prob}). Fact is, what kind of forensic analysis of
the slate could have told me the odds of ``across'' vis-\`a-vis ``along?''  In
all honesty, I could only say

 \begin{quotation}\noindent
 Two roads diverged in a yellow wood,\\
 And I traveled both,\\
 And described to you both of them.\\
 Now \emph{you} choose---I was only a bookkeper!
 \end{quotation}

 \medskip\centerline{*****}\medskip

\begin{dialog}
 A little later that afternoon.
 \medskip

 \cue{You}{Sorry for that smudgy slate! To make up for it, as I was coming back
from the doctor's I typed on my laptop the rule for the system I gave you this
morning---and a new set of initial conditions.  The data are here on this {\sc
usb} memory stick. Can you please run the experiment again?}
 \cue{Me}{Sure! But right now I have to pick up my daughter and stay with her
through soccer practice. I'll be back in three hours.}
 \cue{You}{Fine! I can wait---I got plenty of stuff to do myself.}
 \end{dialog}

 \medskip After the embarrassment of missing my reading glasses, I feel a bit
paranoid. So before leaving I put the memory stick inside a
very-high-resolution particle detector---I've read that ``electronic memory
performance is highly affected by cosmic rays.''

When I come back, I analyze the content of the memory stick and see that the
set of initial nodes reads $\{4,7,9\}$. However, I discover that unfortunately
one high-energy gamma-ray has hit the chip, passing through the
least-significant bit of the memory word where that number 4 is
stored. Typically, if anything happened at all, that bit would have been
\emph{cleared}---\ie for this kind of memory, been forced to the value 0. So
the least-significant bit of that 4 (0100 in binary), which now reads 0, may
originally have been a 0 as now, or may have started out as a 1 and then been
turned into a 0 by the cosmic ray. So, even if that bit had \emph{not} been
changed by the ray, I still don't know whether what now spells 4 was originally
a 4 or a 5. In doubt, I'll have to count \emph{both} 4 and 5 among the possible
initial nodes---in addition to the other three nodes in the given list.\Foot
 {Of course, if a 5 was already present in that list I would have concluded
that the affected node must indeed have been a 4 to begin with, and not added a
second 5 to the list.}

\medskip

Well, you can guess the rest now. The description you wrote in the memory stick
certainly had an initial entropy of 4, because there were only four entries in
your initial-state set. The gamma-ray accident forced me to augment this set to
$\{4,5,7,9\}$---entropy 5.  The law is clearly legible and the system is indeed
\emph{invertible}, so that---barring new accidents---the entropy will remain at
5 forever. However, at some time during my entire accounting process the
entropy has \emph{increased} from 3 to 4 in spite of the system's being
invertible.

 \Sect[2nd]{The second law}

The morale of Day 3's dialogs (\sect{increase}) is that
 \Law[honest]{the honest entropy\\of an invertible dynamical system\\
 never decreases, but it \underline{may increase}.
 }
 \noindent{\sc What}? What happened to the morale of Day 1? (top of
\sect{micro}) Are we in Orwell's {\sl Animal Factory} after the pigs' takeover?
There, the cherished slogan painted on the barn's wall---which every barn
animal remembered as

 \begin{center}
 \sc all animals are equal
 \end{center}

\noindent---was one day found augmented to

 \begin{center} \sc
 \sc all animals are equal\\
 b\,u\raisebox{-1.5pt}{t} s\,om\raisebox{1pt}{e} 
ar\raisebox{1pt}{e} m\raisebox{1pt}\,or\raisebox{1pt}{e} equal \raisebox{-1.5pt}{
t}han \,o\raisebox{-1.5pt}{th}\raisebox{1pt}{e}rs.
 \end{center}

\noindent What has changed? Well, the new morale speaks of ``honest'' entropy,
not of ``internal'' entropy. If we want to turn morale \eq{honest} from a mere
afterthought over the two isolated thought experiments of Day 3's dialogs into
a general \emph{theorem}, we'll have to properly define ``honest entropy.''

\Subsect[contract]{Honest entropy as a contract}

Here we shall first motivate and then define \emph{honest entropy}.  By
proposing such a definition, my goal is to \emph{capture} the most natural (and
plausibly a most generally useful) concept of entropy, if there were such a
thing; if not, to help \emph{shape} one. As we shall see (\sect{decon}), our
definition will naturally extend to the standard entropy context. Our
definition applies not only to the updating of a description by a
\emph{deterministic} law, but also to the case of a \emph{nondeterministic}
law, regardless of whether the state of a system is given as a subset of
ur-states or a probability distribution.

\medskip

The nature of honest entropy reflects a \emph{contract} much like that between
an executive and a consultant. The executive has the \emph{power to decide} how
to use his institution's resources for the institution's greater good, but may
not not be sure of the \emph{consequences} of a given decision. So he decides
to set aside some of those resources to hire a consultant to quickly
\emph{estimate} the actual consequences that a hypothetical decision taken now
will have \emph{in two years}.

The consultant 
 will set up and run at 100\by\ speed (two years to a week) a simplified
version of the world which mimics essential aspects of the problem at hand. For
the reason presented in \sect{prob}, this will typically be a
\emph{probabilistic} model, consisting of (a) an initial probability distribution
over a set of ur-states, and (b) a law describing its evolution in time.
(A lean-entropy model, as indicated in \sect{poormans}, will do as well).

This is fine conceptually, but in practice the ur-states are vastly too many to
be treated one-by-one. In fact the number of ur-states grows
\emph{exponentially} with the system's amount of detail---the number of its
parts or features. 

 \xtra{If my Social Security has ``only'' 9 pieces---its nine digits---that
nonetheless means that the Social Security Administration must be ready to
handle up to a billion files!}
 The ur-state of even a simplified model of something like the weather or the
market may easily run in the billions of digits, alphabetic characters, or
pixels; a liter of gas may have $10^{24}$ parts---and the number of
\emph{ur-states} is an \emph{exponential of that}!  While it may still be
practical to determine the successor of a \emph{single} ur-state and
iteratively follow it along its orbit, a fully general distribution consists of
the weights of all the ur-states, and thus has as much bulk as the entire
ur-state set.  Only distributions that can be represented in a compact way
owing to special properties can be handled, and only evolution rules that can
themselves take advantage of those special properties can be applied; that is,
the restrictions in the nature of a distribution also restrict what the
dynamics itself can express. Thus, one should strive to retain a dynamics that
can easily carry over across an evolution step at least those correlations
(between parts of the model) that are of critical relevance.

In sum, the limited resources available force the consultant to drastically
\emph{condense}, and thus severely \emph{approximate}, both the static
description of the ``state of the world'' (as far as the executive's query is
concerned) and its dynamics. In light of the circumstances, this may still be
an honest manner to proceed on the consultant's part.

\medskip

Within a week, the consultant delivers his forecast for the ``state of the
world'' two years from now---and presents his invoice. How is our executive to
determine if the contract was indeed fulfilled and that the product is worth
its cost?

To be concrete, suppose that one element of the forecast were the market price
of gold---as of two years from now. By its nature, the model will give for this
price a probability distribution, typically approximating a \emph{normal
distribution}, with a certain \emph{mean} (say, \$1,301/oz) and a certain
\emph{standard deviation} or \emph{spread}, say, $\pm$\$700/oz. Now, such a
spread is so large as to make the whole forecast business virtually worthless,
because anyone could have told that ``somewhere between \$600/oz and
\$2,000/oz'' is practically a sure bet---no need to pay a consultant!

Would then our executive accept without a qualm a forecast of \$1,301$\pm$1/oz?
With such an unbelievably narrow spread, he should sue the consultant for fraud
and ask the court to immediately impound all his records: there aren't enough
resources on entire planet to grant such a sharp prediction!

\medskip

From the above discussion one can begin to tease out rational terms of contract
for engaging a forecasting consultant, along the following lines:
 \begin{enumerate}\squeeze
 \item The principal of the contract will have the option to engage an
\emph{auditor} to monitor the performance of the contract, from its onset
(including setting up and running the model) up to one year after the
forecast's object date.
 \item \emph{No limits shall be set by the contractor to the amount of
resources granted to the auditor for his task}.
 \item All the data and procedures used by the contractor during the
performance shall be made available to the auditor, and documented in such a
way as to make the entire contract execution repeatable at will. Rationale and
aims of unusual or ad hoc data-processing procedures should be clearly stated.
 \item The auditor shall not disclose any of the intellectual property
originated by the contractor during the performance, except as evidence of
breach of contract.
 \item The contractor may be fined for any breaches of bookkeeping and/or
accounting rules, as indicated by the auditor, in the following matters:
 \item {\bf Random error.} If, at any stage of the performance, simplifications
or approximations of the model are introduced, or the model is affected by
accidental disturbances, or just \emph{exposed} to such disturbances, a
detailed record of such events and exposures shall be kept. An upper bound
to the entropy gain attendant to any such disturbance shall be provided
\emph{independently} for each event, and accumulated into an external
\emph{entropy burden}, which shall be added to the measurable (by inspection)
``raw entropy'' entropy of the final forecast.
 \end{enumerate}

\noindent At the same time as the auditor is charged with detecting violations
to the above \emph{random error} clause, it might be expedient to additionally
direct his attention to the following \emph{systematic error} clause:

 \begin{enumerate}[resume]\squeeze
 \item {\bf Systematic error}. If, at any stage of the contract performance,
transformations are applied to the model's current state that, in light of the
current state of the modeling art, grossly or unaccountably depart from the
goal of converging toward the desired distribution \emph{mean}, such deviations
\emph{must} be reported and accumulated in a qualitative way into an external
\emph{center-of-mass burden}.
 \end{enumerate}

But, rather than with \emph{honesty}, the latter clause, about systematic
error, has to do with \emph{competence}---which is much harder to quantify. As
Shannon discovered, among parameters that one could use to characterize a
transformation, \emph{entropy change} sticks out because any administrator with
\emph{generic training} can routinely determine it---much as he can determine
the \emph{number} of a candidate's publications simply by \emph{counting} them.
More significant aspects of a transformation---or of a candidate---may require
a great deal of \emph{specific expertise} to evaluate.  Desirable as it may be
in a contract, we shall not include the last desideratum in the contract that
defines honest entropy.

\bigskip

We've seen what the entropy of a description is. We've seen how this entropy
changes as the description is updated by the literal application of a certain
\emph{internal} dynamics. We shall now define the \emph{honest entropy} of a
sequence of transformations.

\smallskip

\noindent {\bf Honest entropy}.\quad Given a ``master'' system that starting
from a certain state has undergone a certain number of tranformations, an
\emph{honest entropy change} is defined as the \emph{computed} entropy change
of an independently simulated version of the system, started in an equivalent
state and subjected to equivalent transformations. This \emph{equivalence}
shall be a mapping of states and transformations of the master system to those
of the simulated system. \emph{For any point in which this mapping is not
one-to-one,\Foot
 {It could be $n$-to-one, as in a homomorphism, or one-to-$n$, as in a Monte
Carlo simulation, in both cases with a lean-entropy change of $n-1$, but with
\emph{different signs} in the two cases.}
 an upper bound to the attendant entropy change shall be assigned and the
\emph{absolute value} of it shall be added to an \emph{entropy burden}
accumulator}. The final value of this entropy burden shall be added to the
difference between final and initial entropies of the simulated system (which
can be determined by inspection).

\medskip

As for the procedural rules stipulated by the above definition, I didn't pull
them out of a hat. They essentially reflect the standard conventions for the
\emph{propagation of uncertainty} (as in \cite{Wikipedia16propagation}, for
one) in experimentation, simulation, and calculation.

I have tried to illustrate the rationale of the above procedural guidelines in
the two dialogs of Day 3 (\sect{increase}), where external events superpose
perturbations, or ``noise,'' on the internal evolution of a system: in one case
on the system's rule and in the other on the system's state.  Even though these
events were \emph{external} to the system proper, honest dealing with them lead
to a \emph{smearing} of our description of the system's state, as compared to
what it would have been by pure internal evolution.

\smallskip

\xtra{For a familiar example of uncertainty propagation, suppose that in a
numerical simulation one \emph{rounds off} all numbers to three significant
figures after each updating step. Then 3.14159 becomes 3.14, but the same value
will have been arrived at from \emph{any number} between $3.135\dots$ and
$3.145\dots$, with an uncertainty range of 0.01. Thus, even though roundoff is
a many-to-one mapping, and so naively may seem to \emph{reduce} uncertainty, in
fact it will have destroyed part of the record, and thus leave us with an
uncertainty range of 0.01 as to the \emph{original content} of the record.

If the original number had been 3.14 to begin with, no uncertainty range---no
entropy burden---need be attached to it because of the roundoff.

Finally, if the same number had been encoded onto an \emph{analog} signal
subject to a noise level of 0.1, and then turned back to \emph{digital} form
as---say, 3.22---by A/D conversion, we should associate with the latter un
uncertainty range of 0.1.}

\medskip

Recall that in ``internal entropy'' the term `internal' does not refer to an
entropy \emph{per se}, but to a particular \emph{discipline} through which an
initial description is successively updated to reflect the effects that a
certain sequence of transformations had on the state of a system. The object of
the game was to give an accurate description of the system's state at the end
of these transformations.

In an analogous way, in ``honest entropy'' the term `honest' does not refer to
an entropy value \emph{per se}, but to a particular \emph{discipline} used to
arrive at it---usually when, as almost always the case---drastic
simplifications are unavoidable.

Even within that discpline, an honest entropy is not \emph{unique}; it depends
on the specific simplification made. More drastic simplifications represent a
coarser approximation of the system, and thus, when honestly acknowledged,
yield a higher honest entropy. The latter only gives an \emph{upper bound} to
the entropy one could have arrived at by making a more accurate simulation of
the master system. The best \emph{lower} bound, of course, is the
\emph{internal} entropy.

\medskip

And what is the \emph{auditor} for? Clearly the uncertainty of a forecast---its
\emph{entropy burden}---decreases its value, and a consultant will be tempted
not to report the full extent of it---as we all may be tempted with our tax
burden. The likelihood of an audit should encourage our consultant to police
himself and not push too hard the envelope\Foot
 {You may be amused to hear that this idiom comes from the mathematics of
curves via the lingo of airplane performance limits---nothing to do with
stationery.}
 of the contract.

\bigskip

Any way of massaging or updating a description---even outright forgery---still
leaves a description of sorts. For this reason, we shall still generally call
simply ``entropy'' the count of all possible instantiations of that
description---as questionable as the latter's provenance may be.

Most forms of entropy that are not obtained via the \emph{internal entropy}
discipline are traditionally called \emph{macroscopic} entropy (of which
\emph{coarse-grained} entropy is a special case). Often a macroscopic entropy
is arrived at by a \emph{hybrid} process. That is, the description is first
updated by a \emph{blind-reckoning} procedure, that is, one \emph{computes} a
few steps of the evolution of the system from given initial data and a given
law, \emph{without access} to the system itself. Then one determines \emph{by
direct measurement} some easily quantified parameter---usually a global
macroscopic parameter such as mass, temperature, pressure, or the results of a
poll; in other words, one ``takes a peek'' at the system. Finally, those two
components are combined into a single description, and the entropy of
\emph{that} description is reported.  An example of this is given in
\sect{ehren}.

 \xtra{The concept of ``macroscopic'' in physics is actually a ``kitchen
sink'' of more-or-less precise intuitions. In Gregg Jaeger's words,
``Macroscopicity turns out to be a rather vaguer and less consistently
understood notion than typically assumed by physicists who have not explicitly
explored the notion themselves.''\cite{Jaeger14macro} Also, in
\cite{Jaeger14quantum}, ``Long before the appearance of the term in quantum
theory, consistent standard but different use of the notion of the macroscopic
was made in thermal physics and statistical mechanics, in reference to the
macroscopic variable and in connection with the relationship of the statistics
of basic components to thermodynamical variables, \eg temperature and pressure
of collections.''}

\medskip

The concept of \emph{honest entropy} may provide a quality-control reference
point (``here we know what we are talking about'') in the commerce of
macroscopic entropy.

\bigskip

What we shall do next, is present in a discursive way, always in the context of
the present lean-entropy approach (``entropy-as-a-count,'' as defined in
\sect{entropy001}) an important conclusion of the present paper, concerning the
relation between \emph{honest entropy} and the \emph{second law of
thermodynamics}. We shall briefly revisit the second law after the
deconstruction/reconstruction excursus of \sect{decon}.

\medskip

Let us first recall that the \emph{second law of thermodynamics}, as formulated
by Clausius on empirical grounds, can be paraphrased as
 \Law[clausius]{the entropy of an isolated physical system\\ never decreases,
 }
 \noindent and that it is believed that
 \Law[physinv]{the laws of physics are at bottom\\strictly invertible.
 }
 \noindent Further recall, from theorem \eq{internal}, that
 \Law[inv-const]{the internal entropy\\of an invertible system\\is constant
 }
 \noindent(note that the object of internal-entropy bookkeping is always an
\emph{isolated} system), while from \sect{decrease} we have (as per
\foot{proof}) the following theorem:
 \Law[th-in-noninv]{the internal entropy\\of a noninvertible system\\
  never increases\\
  but occasionally does decrease.
 }

\Subsect[weak]{The weak second law}

Let us now run through a trivial exercise of logic. From theorems
\eq{inv-const} and \eq{th-in-noninv} we obtain a \emph{weak second law},
namely, that
 \Law[th-iff-weak]{a system's internal entropy is constant\\ if and only
if it is invertible.
 }
 From that law and assumption \eq{physinv} we get, as a corollary, that
 \Law[th-phys2nd]{if a system is physical\\it must obey the weak second law\\
  (that is, its internal entropy is constant!)
 }

 \noindent We shall as well explicitly state the converse of theorem
\eq{th-iff-weak}, namely,
 \Law[th-converse]{should its internal entropy ever decrease,\\
  the system violates the weak second law\\
  (and therefore must be noninvertible!)
 }

\Subsect[strong]{The strong second law (important!)}

Using an argument formally similar to that used above for the weak law, but
starting from a more practically interesting premise, I'll arrive at a much
stronger conclusion.

What I proclaim, and will just for the moment call the \emph{strong second law
of thermodynamics}, is none but
 \Law[law-honest-monot]{a dynamical system's honest entropy\\
 never decreases\\if and only if the system is invertible!
 }

\noindent The proof is is straightforward; it follows directly from the
definition of honest entropy in \sect{contract} with the help of the following
table
 {\small
 \Eq[budget]{
  \def\strut{\vrule height12pt depth6pt width0pt}
 \vcenter{\hbox{
 \renewcommand{\arraystretch}{1}
 \begin{tabular}{c|c|c|}
  \multicolumn{3}{c}{\sc Honest-entropy change through a step}\\
  \multicolumn{1}{c}{} & \multicolumn{1}{c}{noninvertible} &
			 \multicolumn{1}{c}{invertible}\\
  \multicolumn{1}{c}{} & \multicolumn{1}{c}{(many-to-one)} &
			 \multicolumn{1}{c}{(one-to-one)} \\\cline{2-3}
  \strut internal & $\leq0^{(*)}$ & $=0$ \\\cline{2-3}
  \strut external noise	& $\geq0$ & $\geq0$ \\\cline{2-3}
  \multicolumn{1}{c}{\strut {\sc total}}	&
  \multicolumn{1}{c}{$\lesseqqgtr0$}		&
  \multicolumn{1}{c}{$\geq0$}			\\	
  \multicolumn{3}{c}{\footnotesize $(^*)$ Strictly less for at least one state
transition}\\
 \end{tabular}
 }}}}
 \par\noindent As per this table, at any step the system's honest
entropy will
 \begin{itemize}\squeeze
 \item if the system is \emph{invertible}: remain constant, or increase;
 \item if the system is \emph{noninvertible}: decrease, remain constant, or
increase;  moreover,
 \item if for some appropriate current state description and low enough
external noise a \emph{decrease} can occur \emph{at all}, then the system is
\emph{noninvertible}.\quad {\sc qed}
 \end{itemize}


Note the \emph{asymmetry} between the {\sc if} and the {\sc only if}. I can
check that a system is invertible by its law, which is given to me, and from
that I can make nonvacuous predictions about its entropy changes in different
circumstances (noise vs no noise). Similarly, I can check that a system is
noninvertible, by its law, and from that make some nonvacuous predictions about
its entropy changes. 

For the converse, however---that is, if I want to infer something about the
invertibility of the system from the changes in entropy---the task is made more
difficult by the presence of external noise. If I \emph{happen to} hit a
\emph{single} state transition for which the entropy \emph{decreases}, then,
irrespective of external noise, I can conclude that the system is
noninvertible. However, even if I manage to turn off all noise, to conclude
that the system is \emph{invertible} merely from its entropy changes I have to
test \emph{all} possible state transitions. In the limit as the system's size
goes to infinity, the experimental determination of invertibility vs
noninvertibility on the basis of observed entropy changes (we are still
speaking of entropy \emph{of a description}) become de facto only
\emph{semi}-decidable.

\bigskip

Our strong second law has a nice feature. It reconciles Clausius's sweeping
statement, that ``the entropy of the world \emph{increases},'' based on
empirical thermodynamics near equilibrium, with the statistical mechanical
interpretation---which in certain ``non-honest'' conditions makes room for
entropy fluctuations.  We owe this reconciliation to the
humble \emph{honest entropy}.

A worked-out example of reconciliation of the paradox---up-and-down
fluctuations vs monotonical increase of an expected value---is given in
\sect{ehren}. There, the score of a game is updated in real-time, while a
forecaster updates, also in real-time, an \emph{expected} score based
\emph{only} (a) on information available \emph{at the beginning of the game}
and (b) just how many rounds of the game have been played so far---but none of
the actual outcomes. In the actual game, the score may well go up and down; in
these ``blind-reckoning'' forecasts, the expected score will monotonically
increase \emph{if and only if} the system is \emph{invertible}!

\medskip

On reflection, all this should not be surprising. Honest entropy is but an
expression of the MaxEnt principle (see \sect{entropy000}, Jaynes entry) in the
context of a \emph{forecasting} context (where by necessity one must perforce
\emph{anticipate} the state of a system without seeing it, as the contractor of
\sect{contract}), or whenever such ``blind reckoning'' is stipulated by the
rules of engagement.

And this is the source of the generality of the strong second law of
thermodynamics, and by implication, of Clausius's second law!

\Subsect[not]{Not a \underline{physical} law?}

In the title of this section (see Myth 7 in the Introduction), remark that what
is emphasized is not ``law'' or ``not,'' but just the word ``physical.'' Now I
shall explain what I mean; to the cognoscenti of you, this will not come as a
surprise. The surprise, if any, may be that---as in Andersen's ``The
Emperor's Clothes''---somebody had the naivet\'e to say it aloud!

\xtra{Every morning a shepherd would take his sheep out of the fold and every
night he would put them back in. Sheep, like kindergarden children, have a way
of getting out of line for any distraction. So the shepherd had to count and
recount his sheep to make sure he got them all in. This shepherd had an
inquisitive mind---and plenty of time to think. One day he observed, ``Isn't
this an interesting coincidence? When I put seven sheep in the fold, and then
three more, I end up with the same number of sheep as when I start with three
and then add seven.'' Our ``protoscientist'' experiments with different
combinations of sheep numbers, and such a fact always holds true. ``I have
discovered a great physical law,'' he concludes one day, ``I'll call it the
\emph{Second Law of Sheepdynamics}!''}

I grant the shepherd the independent discovery of a law. But to call it a
\emph{physical} law? I agree that this law is correct.  Indeed, few things are
truer than it---and that's where the catch is! By the same token, the
cheesemaker will (correctly) claim that cheese counting obeys a Second Law of
\emph{Cheesemaking}, and similarly the hot-chestnut seller, a law of
\emph{Chestnutroasting}. This is of course harmless, but a bit parochial: will
you call those a \emph{dairy} law and a \emph{roasting} law?  They are all
already subsumed under a \emph{Commutative Law of Addition}, and, to paraphrase
Ockham, in science it is just \emph{not good manners} to introduce more laws
than necessary.

The second law of thermodynamics is at bottom a \emph{counting inequality},
valid, as we've seen in \sect{strong}, for all and only \emph{invertible
dynamical systems}---be they physical as a gas or abstract as a Turing machine
or a cellular automaton. It can be thought of as a law of \emph{category
theory}.

As a \emph{corollary} of that general law, the second law becomes a
``\emph{physical} law'' on the mere premise that physics be at bottom an
\emph{invertible} dynamical system.  In other words, as soon as one proposes
for physics (say, on the basis of overwhelming experimental evidence) a
\emph{model} that is at bottom an \emph{invertible} dynamics, physics
automatically inherits the second law as an intrinsic property of that model.

\xtra{In spite of my irreverence, I'm emotionally very attached to the
invertible dynamics model of our universe, and I'd place a large bet on
it. However, just to play devil's advocate, one may argue that the observed
invertibility of microphysics might be just an \emph{emergent property} of a
\emph{noninvertible} substrate. For example, in a \emph{finite} (though no
matter how large) deterministic dynamical system, any trajectory eventually
falls into the attractor of (or ``merges with'') a closed orbit. The collection
of these closed orbits is of course an \emph{invertible} system. So, after the
system has run for a long time and most of the merges have occurred (to
paraphrase Feyman's, ``When most fast things have happened'') and the remaining
merges are ever rarer and farther between, the system will be \emph{de facto}
invertible and thus \emph{de facto} obey the second law (eventually it will
obey it strictly) in spite of its defining dynamics being
\emph{noninvertible}. I briefly discussed this devil's advocate's
counterexample with Feynman in 1981, and he had no objections to the
counterexample per se; the real question was of course whether such an ``aged''
system would still have enough stamina to support, on top and in spite of the
aforementioned emergence of invertibility, complex emergent phenomena like
life.}

\medskip

Physicists deserve praise for having been first at \emph{identifying} and
\emph{formulating} the second law of thermodynamics. After Boltzmann, however,
it must have been clear that this is a law of enormous generality, and thus
belongs to \emph{everybody}, not only physicists. It has become \emph{public
domain}!

\xtra{Incidentally, this law should properly be called a law of
thermo\emph{statics}, since it does not prescribe a motion, but only forbids
motion in certain \emph{directions}. In physics, in a course of \emph{Statics},
one studies when a scale is at equilibrium. But the principles of statics have
something to say even when a system is \emph{not} at equilibrium. Namely, they
will tell us to which side the scale will \emph{not} tip---though not whether,
when, or how fast it \emph{will} tip to the other side.}

\noindent Specifically, the second law of thermodynamics is valid for ``any
invertible dynamical system from which one is temporarily cut off.''  In this
sense, it is a \emph{law of logic} in the form of information theory.  Whenever
you forget a bit about a system's state, or a tooth slips in your mechanical
simulation of it, you are in effect erasing one \emph{constraint} from the
state's current description. By the MaxEnt principle, honesty or self-interest
require that, every time you remove a constraint, you let the distribution that
embodies that description spread out to ``fill the slack'' and thus remain
MaxEnt.

It turns out that physics \emph{does} meet the above description. But \emph{a
countless variety of other systems do too}. On the other end, if the system is
not as defined above---or it does, but at a certain point you restore your
contact with it----then this law no longer need apply (see \sect{ehren} for an
example). It is in this sense that I argue that the second law is not a law of
physics \emph{per se} (such as instead, I suppose, is the first
law---``conservation of energy''), much as the commutation property of sheep is
not a law of shepherding \emph{per se}.

\bigskip

The second law's intuitive explanation is that we---a liter of gas, the
weather, etc.---are localized patches of a large, \emph{distributed} (roughly,
``spread out'' rather than ``lumped''), yet \emph{interconnected} system. In
this situation \emph{no agency} can have \emph{total control} of any portion of
the universe---and fortunately so!  Mice and men may claim a patch of this
universe as exclusively their own, but their best-laid plans can never manage
to unconditionally isolate and protect it from external influences. The latter
may be felt by the claimants as ``random disturbances'' (that is, ``hard to
predict''), even when they may be a consequence (intended or unintended) of
what other mice and men (or galactic agencies) are quite deliberately and
cognizantly (thus nonrandomly from their viewpoint) doing with \emph{their}
patches. Or they may simply be a minor accidental side-effect of the explosion
of a dumb supernova!

In the above paragraph we didn't explicitly mention either physics or
invertibility; we just envisioned any large distributed system with enough
\emph{coupling} (or ``interaction between parts'') to keep it exciting---as
long as that lasts.  Politics, economics, and ecology are good examples.

You may object that at that level of aggregation the dynamics of a system may
no longer be seen as invertible---even though that of the physical substrate
were, and therefore there wouldn't any longer be a guarantee that the second
principle strictly held. That is true, but if you look at the right column
(\emph{noninvertible}) of table \eq{budget}, you will see that the net entropy
change would remain $\geq0$ as long as the ``forking of paths''
(\sect{increase}) contributed by the \emph{external} noise---the crowds
jostling about you---overwhelmed the the ``merging of paths''
(\sect{decrease})---dictated by a noninvertible internal dynamics. The above
state of affairs is a very likely one in almost any conceivable universe,
invertible or not, considering that, if the ``noninvertibility'' of a system
were too drastic, everything would soon ``freeze'' and there wouldn't be any
one left to ask questions. According to Norman Margolus (personal
communication), an \emph{invertibile} pinball machine was the Creator's obvious
choice, for it yields ``the most playing time for your quarter!''

\Subsect[ehren]{Fluctuations paradoxes; macroscopic entropy}

Since its introduction, the second law has been (and still is) plagued by
objections and paradoxes---chief representatives of which are the
\emph{Loschmidt} (or \emph{reversal}) paradox (Kelvin 1874, Loschmidt 1876),
the \emph{recurrence} paradox (Zermelo 1896), and the well known
\emph{Maxwell's demon} paradox (Maxwell 1866).

In short, these paradoxes present theoretical or empirical examples where in
the course of time physical entropy would spontaneously \emph{decrease}, in
violation of the second law. At the core of these paradoxes there usually is
some version of the \emph{mind projection fallacy} (see \sect{prob}) where one
starts with a mental construct such as ``witchness,'' and incorrectly assumes
that if there is a ``property of being a witch'' there must exist an object
with that property---and happily goes on a witch hunt.

\medskip

We've already given two toy examples of this---the witch hunt for \emph{random}
numbers in a list of integers (\sect{myths}), and the ``concrete example'' in
\sect{entropy001}. In the latter, the high entropy of a shuffled deck
miraculously collapses---by just looking at the deck---to the the minimal entropy of
a virgin deck (1 in our count scale, 0 in the conventional log-count scale). No
amount of torture will extract from that deck a confession of \emph{its}
current entropy, since there is no such thing \emph{in the deck itself}.

Here we shall give, as a more detailed example---a streamlined version of that
proposed by Paul and Tatiana Ehrenfest as representative of the
\emph{fluctuations} paradox\cite{Ehrenfest07objections}.

Consider a row of 100 checker pieces, or \emph{tokens}, but white on one face
and black on the other, and a bingo machine from which at every shot one will
obtain one at random of 100 balls numbered 1--100, to be announced and
immediately placed back in the basket. An ur-state of the row is any specific
configuration of face \emph{colors} (black and white) for the tokens; there are
thus $2^{100}$ different ur-states.

Initially, all 100 tokens show white faces. This information uniquely
characterizes the system's state, and so this state's entropy is 1. Balls are
drawn sequentially, and, every time ball $n$ is drawn, token $n$ is turned over
(or ``flipped''). The row will display an increasing number of black tokens,
and soon some black tokens will start to turn white again. When black and white
tokens are close to an equal number, the frequency of black$\to$white
transitions will approximately match that of white$\to$black ones, and the
white:black ratio will be dynamically maintained at approximately 50:50, where
the entropy is maximal. While, as we've seen, there is only one configurations
showing no black tokens, there are $\approx10^{29}$ configurations with a 50:50
population!  Small population fluctuations about the 50:50 ratio are common,
and thus corresponding departures of entropy from that ceiling value of
$\approx10^{29}$, or ``entropy fluctuations,'' are common.

\medskip

But on what conditions? Suppose that instead of 100 tokens in a row we had a
square of one-million by one-million tokens (a trillion in all) displayed on a
screen. For comparison, an ordinary TV screen has only about 1000\by1000
pixels; this number is chosen to be large enough that one can barely make out
the individual pixels. Then in practice, when viewing our entire
1,000,000\by1,000,000 screenful of tokens, we won't be able to distinguish
configurations that differ just by the color of a few individual tokens, since
from that distance we'll have a slightly blurred picture!  And the value---in
terms of information conveyed---of a blurred picture is inversely proportional
to the amount of blurring---that is, the \emph{number} of different
fine-grained pictures that in effect have been confused with one another by the
blurring. This is indeed the \emph{entropy}---or, with Clausius original term
(\cf \sect{entropy000}), the ``transformation contents'' of the blurring
transformation.

In spite of the blurring, all is not lost. The overall intensity of black on
white scattered over a surface is \emph{additive}. In half-screen
printing---that fine mesh of variable-size dots barely perceptible in the
texture of a newspaper picture---a number of \emph{black} dots of a certain
size over an otherwise \emph{white} area, collectively give the visual
impression of a certain \emph{level of grey}.  And \emph{that level} is
something we can easily perceive, or measure with a photocell; this will
provide an indirect way to estimate at least the number, if not the precise
position, of black tokens present in a certain area.

As a result, we do receive some information, and the entropy of the picture we
see is less than that of knowing that ``it could be just any picture.'' In
effect we are allowed to \emph{peek}---though through blurred glasses.

\smallskip

Returning to our row of tokens, let's suppose that the whole updating process
is taking place in a separate room, out of my sight and hearing. There tokens
are flipped according to ``bingo calls'' like ``Flip token 7!,'' ``Flip token
75!,'' etc., regularly spaced one second apart. Suppose further that after a
while I am invited to bet a dollar on the current token configuration, with
odds 1:$n$---that is, I get $n$ dollars if I've hit on the right
configuration. What value of $n$ should I insist on for the game to be fair?

With  this picture in mind, let's consider the following three scenarios:
 \begin{description}
 \item[internal entropy] (a) I'm given the initial token configuration, (b) I
can hear the bingo calls through an intercom, and (c) I am willing and capable
to update my mental model of the process according to the information given to
me.

Then anything above one dollar will be reward enough, since I'll be able to
track the process in \emph{full detail} and \emph{on my own}. That will make mine
a sure bet! No needs to bother entropy.

\xtra{Remark that the law of this dynamical system---the entire sequence of
flips announced to me---is \emph{deterministic} and \emph{invertible}. To see
this, suppose that my friend Elsie wants to ``predict'' the whole sequence of
token configuration \emph{backwards}, starting from the \emph{last}
configuration, using only the laws of the dynamics that had been applied to go
forward. To this purpose I give her the sequence of bingo calls as it had been
given to me; this sequence, together with the provision that each called token
is to be flipped, represent the entire law. Starting with the \emph{last}
configuration as an initial state, she applies that sequence of instructions,
but using the \emph{inverse} operations in the \emph{reverse} order. Now, the
inverse of ``flip'' happens to be again ``flip,'' so Elsie will progressively
unflip each token that had been flipped, finally holding in her hand the
\emph{initial} configuration as reconstructed from the \emph{last} and the
\emph{law}.}

 \item[honest entropy with an external component] In this scenario, I cannot
hear the \emph{contents} of the bingo calls---but I can still infer \emph{when}
each is made by looking at my watch.

In this case, after one step the choice of possible configurations jumps from 1
to 100; in fact, I know \emph{that} one token has flipped, but I don't know
\emph{which}. Note that the law is \emph{deterministic} for the guy who flips
the tokens in the other room---it is being ``dictated'' to him by the bingo
caller and thus he has no choice. However it is \emph{nondeterministic} for
\emph{me}, as I don't hear the numbers called.

\xtra{On the \emph{second} step, in one case out of 100 (but I don't know
which) the caller will draw the same number as on the first step, in which case
that token will get unflipped and we go back to the original configuration---a
single ur-state. In the other 99 out of 100 cases a \emph{new} token will be
hit, so the result will be \emph{two} black-faced token---and of distinct
configurations like that there are 100\by99/2. By taking the geometric mean of
the two situations, weighted respectively 1 and 99, we get an equivalent
ur-state count of $(99\cdot50)^{99/100}\approx4546$. And so forth, step after
step of blind reckoning with more and more complicated expressions for the
expected equivalent count, converging to an asymptotic value of
$\binom{100}{50}$ for $t\to\infty$, corresponding to an expected number of 50
black tokens.}

Note that even though the actual number of black tokens may go up and down, the
above sequence of entropy values, relating to my state of knowledge, is
strictly \emph{monotonically increasing}, with no fluctuations whatsoever.

 \item[partial peeking] In this last scenario, I do not want to do all
those tedious computations to determine the extent of my
uncertainty. Fortunately I discover that in the wall between the two rooms
there is a groundglass window. The blurry view through it does not let me
distinguish the individual tokens, but it lets me estimate the number of black
tokens by the level of gray of the row as a whole seen as a blurry patch. Then,
by monitoring this number \emph{in real time} (as ``parasite'' of the actual
game being played in the next room, I can enjoy some of it without having to
perform a simulation of it myself), I can infer the number of token
configurations corresponding to the current estimated number of black tokens.
Depending on the specific tokens affected step by step, this number is quite
likely to display fluctuations about the general trend, in violation of the
strong second law!
 \end{description}

The entropy obtained in the third scenario is not, of course, a form of what we
called ``honest entropy;'' here I peeked! It is an example of what is commonly
called \emph{coarse-grained} entropy to distinguish it from \emph{microscopic}
entropy (which, as already mentioned, roughly corresponds to what we called
\emph{internal} entropy). Also see the end of \sect{contract}.

\Sect[decon]{Shannon's formula deconstructed} 

\myquote[Andrew Marvell, ``To his coy mistress,'' ca.~1655]
{Had we but world enough, and time,\\
This coyness, Lady, were no crime}

 Here we shall start with the standard entropy formula in Shannon's form
(\eq{shannon}),
 \Eq[std]{S=-\sum_i p_i\ln p_i,
 }
  which, as we've seen, drops the ``physical legacy'' dimensional coefficient
$k$ of \eq{gibbs}, and will deconstruct it in a number of steps in order to subject
it to ``structural analysis.''

 \noindent Though nice and compact, the Shannon formula for entropy is not very
transparent to the uninitiated.
 \begin{enumerate}\squeeze
 \item Why the leading minus sign, since probabilities are positive?
 \item Where does the logarithm come from?
 \item How come the quantity $p_i$ appears two times in the formula? Does it
 play different roles in the two places?
 \item What is all this probability stuff?
 \end{enumerate}
\noindent Finally,
 \begin{enumerate}[resume]
 \item What does the whole thing mean?
 \end{enumerate}

Among other things, we'll argue that, after peeling off the convenient but
inessential logarithmic wrapping, this formula is a generalization of a
\emph{count}, and as such is drop-in ready, as it were, for replacement of the
entropy-as-a-count used in the arguments of the preceding sections.

At the end of the day, we'll conclude that the structure of that formula was
sound---though a bit baroque---and that most of the pieces that we'd taken out
to examine can go right back in.

\medskip

As a first thing, the naive may wonder where that minus sign might have come
from, since one is dealing with positive quantities throughout. Let us
acknowledge that this is a purely \emph{graphical} convenience. After passing
through the logarithmic filter, division become \emph{subtraction} and inverse
become \emph{opposite} (or ``the `negative','' in casual talk).
 When $\log(1/p)$ is written in display style it appears as $\displaystyle \log
\frac1p $, and thus is ugly and wastes vertical space---even more when
we substitute for $p$ a taller expression.  Thus the minus sign does not
indicate something conceptual that you may have missed, but is merely a
spacesaving device. If we waive this ``convenience,'' \eq{std} becomes
 \Eq[std1]{S=\sum_i p_i \ln\frac1{p_i},
 }
 with no minus sign! At this point, one is immediately reminded that, if
$p=\{p_1,\dots,p_n\}$ is a probability distribution and $x = \{x_1,\dots,x_n\}$
a random variable over $i$ (a more modern term is ``random \emph{function}'')
then $\sum_i p_i x_i$ is the \emph{expectation} (or \emph{expected value}) of
$x$, denoted $\text{Exp}_p[x]$.  In our case, \eq{std1} can be rewritten as
 {\small
 \Eq[std2]{S = \text{Exp}_p\left[\ln\frac1p\right],\ \ \text{where}\
\ln\frac1p\ \text{stands for}\
\textstyle\{\ln\frac1{p_1},\dots,\ln\frac1{p_n}\}.
 }}
 The argument of the expectation is, as expected (no pun intended), a random
variable, spelled in a customary shorthand; the full spelling is the expression
in curly braces. The value of the expectation is the \emph{arithmetic average}
of (the values of) the random variable in the argument.  We'll next discuss the
meaning of that random variable.

\Subsect[equi]{Equivalent count}

We've seen in that in \eq{std2}, while $p$ is a \emph{probability
distribution}, $\ln\frac1p$ is instead a \emph{random variable} over the same
set of outcomes (the latter labeled $\{1,2,\dots,n\}$). With the same
shorthand, $1/p$ is also a random variable; the fact that its \emph{value}
$1/p_i$ for outcome $i$ happens to equal the \emph{inverse} of the \emph{value}
$p_i$ of the \emph{probability weight} of the same outcome, is not a
mathematical tautology (like ``$1/x$ is course the inverse of $x$---\emph{by
definition}'') but a \emph{coincidence}---an artifact arising \emph{by
construction} in making up that particular random variable.

\xtra{Imagine an urn containing $n$ marbles which may have different colors,
red, green, blue, etc., and count the number of marbles of each color,
$n_\text{red},n_\text{green},\dots$. Not being interested in the total number
$n$ but only in the \emph{fraction} of marbles of each color, we rewrite that
fraction as a probability distribution $\{p_\text{red},p_\text{green},\dots\}$,
where $p_i=n_i/n$. The value $n$ is thereby lost.

At this point we assign to the marbles \emph{a posteriori} a new
property---let's call it \emph{equivalent count}---besides their color. Remark
that this property will not depend on the physical features of a marble
\emph{per se}---color, radius, mass, or even where it is in the urn---but only
on \emph{how many} marbles of the \emph{same} color there are in the urn
relative to the total number. As a property ``of a marble'' it has a peculiar
behavior. For example, it will change by the simple fact of adding a new
marble, of any color, to the urn, without touching the original one. Even if,
as part of a nationwide lottery game, the ``urn'' consisted of the joint
contents of two widely separate containers, yet making a minimal change to the
contents of one container will affect the equivalent count of a marble in the
other container, even though \emph{no physical interaction} between the two containers has taken place.

In our case, that ``spooky action at a distance,'' as Einstein might have
called it (\cf \cite{Einstein35epr}), has to do with the \emph{self-referential
nature} of the construction of the ``equivalent count'' random function.  In
effect, the latter is a ``property'' of the physical marble only insofar as it
is a property of the \emph{entire distribution}, which in turn is an expresson
of the joint makeup of the two containers. (We had the same issue with the prom
photo example of \sect{entropy001}.) This ``entanglement,'' as it were, takes
place not in the concrete world, but only within the abstract construct of a
distribution---which \emph{by definition} changes \emph{as a whole} whenever
the makeup of the urn is altered.}

\noindent If the $p_i$'s happen to be all equal, then $p\equiv1/n$, where $n$
is the number of outcomes in the distribution of which the $p_i$ are the
weights. Intuitively, if the size of a pizza slice is \emph{one-third} of the
whole pizza \emph{and} the pizza has been cut into \emph{equal} parts, then the
number of slices must be \emph{three}. We can think of the equivalent count of
a single slice of any size (independent of the sizes of the other slices) as
the number of slices \emph{of that size} which could be cut out of the whole
pizza, counting in that number also the fraction of a slice that might be left
as a ``remainder. Thus if we have just two slices, of sizes respectively 1/3
and 2/3, the equivalent count of the first slice will be 3 and that of the
second 1.5 (that is, we can only get one-and-a-half slice of that size from the
whole pizza).

It is clear that $p$ and $1/p$, while numerically just the inverse of one
another, stand for quantities that are of a rather different nature. To
highlight that, it will be helpful to give the random variable \emph{equivalent
count}, born $1/p$, its own symbol $v$; that is, $v_i=1/p_i$.

\medskip

\emph{I've come to believe that much of the confusion, uneasiness, and
controversy surrounding entropy hinges on that disturbing self-referential
nature of ``equivalent count.''}

\xtra{The magic allure of self-reference has been extensively and gloriously
illuminated by logician and magician Raymond Sullyvan (my vote for a ``{\sc
Unesco} World Heritage Unique Human Resource'') in his popular books---for one,
\emph{Forever Undecided: A puzzle guide to G\"odel}\cite{Smullyan87forever}.}

\Subsect[log]{Why logarithms?}
 
From the viewpont of the above subsection, the entropy $S$ of the urn is the
log of the \emph{mean equivalent count} of a certain kind of outcome---in our
case, marble \emph{color} (the mean being taken over all individual marbles).

\smallskip

What happens if we strip the log from $S$ by taking the exponential of the
whole expression---thus obtaining the quantity $V=e^S$? Use of the logarithm is
very convenient in practice (and, as we stress below, also has an important
conceptual role in the physics of information), but is mathematically
irrelevant. It just lowers by 1 the rank of arithmetical operations: a product
turns into a sum, a quotient into a difference, a power into a multiple, and,
as we've seen, an inverse into an opposite.  What's important now is that a
\emph{geometric} mean turns into an \emph{arithmetic} mean.  If we ``undo'' the
log in $S$ we get back the geometric mean. Thus $V$ can be rewritten as follows
(where we use the variable $v$ for the equivalent count $1/p$ as introduced at
the end of \sect{equi}):
 \Eq[up]{
  \textstyle V = e^S = \prod_i {v_i}^{p_i} ={} _\text{G}\text{Exp}[v],}
 where the pre-subscript G reminds one that what we intend is the
\emph{geometric} mean.

We shall call \emph{variety} the quantity $V$ introduced above. Then \eq{up}
reads as ``The variety of a distribution is the (geometric) mean of its
equivalent count.'' If we went whole hog and used \emph{self-variety} as a term
for the equivalent count $v$, then we'd notice the tautological equivalence of
the above sentence with a better known sentence, namely, ``The information (or
entropy) of a distribution is the (arithmetic) mean of its self-information
$\ln\frac1p$.''

\medskip

The term ``variety'' in this general context was introduced by Ross Ashby in
1956\cite{Ashby56variety}. Even though his use of the term with the meaning of
``simple count'' rather than ``equivalent count'' is today deprecated (see
\sect{prob}), he clearly makes the point that reporting a count as such or
through a log of it is a practical rather than conceptual issue:
 \begin{quote}
Variety can be stated as an integer, \dots\ or as the
\emph{logarithm} to the base 2 of the number, \ie in bits.
 \end{quote}

\medskip

More recently, and by wide consensus, \emph{variety} has been
reintroduced---with the present meaning of ``mean equivalent count''---as one
of the most useful indices of diversity\cite{Heip98diverse,Hanel14}.  This mostly in
biology, taxonomy, paleontology, ecology, economics, cataloging and
inventorying, etc. To quote from Straathof's abstract\cite{Straathof03variety},

 \begin{quote}
     The antilog of Shannon's entropy is a suitable index of product variety
for three reasons. First, for symmetric product types it is equal to the number
of product types. Second, disaggregation of the underlying product set always
leads to an increase in measured product variety. Third, the introduction or
disappearance of a marginal type does not cause a discrete change in the
variety index. These properties hold for a class of weights that includes, but
is not limited to, frequencies.
 \end{quote}

The preference in those areas for this ``antilog of Shannon's entropy'' rather
than Shannon's entropy itself---in other words, for dealing with an equivalent
count rather than the log of this count---seems obvious to me. In those
applications, one has to materially \emph{go} through those things that are
counted (the items of an inventory, the species of a certain genus, etc.), and
the resulting number is on a human (or computer) scale, because one is counting
tangible \emph{objects}---whether ``bodies,'' pieces of a data structure, or
computational events.

\smallskip

In physics and in information science, on the other hand, one is often counting
\emph{states} of a system, rather than \emph{pieces} of it, and that grows
\emph{exponentially} with the number of pieces. For a mole of gas, with
$\approx 10^{24}$ molecules, in ordinary conditions one can have some
$\approx10^{10^{24}}$ states!  Physicists can touch with hand a specific
ur-state---and, even that, as a norm, for just a fleeting moment: water
molecules collide a trillion times per second!  As the Greek philosopher
Heraclitus ($\approx470$\,{\sc bc}) put it, ``You cannot step twice into the
same river.''  One can deal with the entire collection of \emph{ur-states}
through a symbolic object---a formula or a numeral.  For example, the above
number's decimal representation, which is already a logarithmic compression of
a tally count (the scrawled daily marks on the wall of a prisoner's cell) is
``only'' a trillion trillion digits, thus large but not astronomical. But one
can never see in one's laboratory or explicitly represent in one's computer all
those $10^\text{trillion trillion}$ \emph{states}, simultaneously or even just
sequentially. This number is \emph{hyper}-astronomical: for comparison the
total number of elementary particles in the universe is estimated to be a
``puny''
 $10^{85}$.

To paraphrase the quote at the beginning of this section, the physicist could
well tell the catalog librarian,
 \begin{quote}
 Had we but world enough, and time,\\
 your spurning of logarithms, Lady, were no crime!
 \end{quote}

Beside the practicality of Marvel's entreats to his coy mistress (the poem's
point is that when he'd be finished counting her beauties they'd both be too
old to be able to enjoy them), there is a strong physical reason for using a
logarithm when counting states of a distributed system. This log essentially
counts the number of \emph{material parts} of the system---which are all
physically present at once. It measures the amount of the \emph{physical
resource} available to generate states with---the quantity of
\emph{state-making stuff}.  On the other hand, the state set---the collection
of \emph{all possible states} of an extended systems is clearly only a
potentiality---a mental fiction. As we've seen, not even a microgram of matter
will be able to go through all its possible quantum states during the age of
the universe, and at any moment will be in only one of them.

\medskip

We've seen that, in its standard form, entropy uses a logarithmic scale just as
a convenient wrapper for the more intuitive \emph{variety} (our ``mean
equivalent count''), and that the leading minus sign in its formula is
graphical space-saving device made available by the presence of the log.

Moreover, in any large multicomponent system---be it a computer or a microgram
of matter---even a tiny part of it won't be able to go through even a tiny
fraction of its state set before the least interaction with the rest of the
world makes its current state, its state set, and its very dynamics obsolete.

\xtra{That is illustrated by Borel's memorable example\cite{Borel12Sirius}, of
how a tiny displacement of a tiny mass on Sirius will totally disrupt, in a few
collisions' time, the trajectories of gas molecules in a sample on Earth.}

We've also seen that, when we partition a collection of marbles by, say, color,
and look at the resulting color classes, a more informative measure of the
collection's variety than merely counting the classes is indeed their
\emph{mean equivalent count}. Giving weight to classes is generalization that
one ``cannot refuse!''

In the next subsection we briefly deconstruct probability.

 \Subsect[prob]{What's probability?}

In Gibbs's and Shannon's formulas above, the letter $p$, which we used for a
generic ``share'' or ``weight,'' actually stands for \emph{probability}, and a
collection $\{p_1,p_2,\dots,p_n\}$ of $p_i$'s with $p_1+p_2+\dots+p_n=1$ and
$p_i\geq0$ is accordingly called a \emph{probability distribution}. 

\medskip

But what is probability? A serious problem was raised by Bruno de
Finetti---economist, mathematician, and a pioneer of neo-Bayesianism.
 We shall see that in phrases like ``this face of the die has probability
1/6''---as well as ``this brick has entropy 18'' or ``7 is a random
number''---do not literally mean that ``having probability 1/6,'' ``entropy
18,'' or ``being random'' are intrinsic properties of the named objects. They
are at best shorthands (and as such often useful) for properties of certain
viewpoints or decisions of an \emph{external} nature. It is in this vein that
on the first day of his probability lectures Prof.~de Finetti would shock his
students with the wake-up call

 \smallskip
 \centerline{\sc probability does not exist!}
 \smallskip
 \noindent And he'd go on:
 \begin{quote} The abandonment of superstitious beliefs about the
existence of Phlogiston, the Cosmic Ether, Absolute Space and Time, \dots, or
Fairies and Witches, was an essential step along the road to scientific
thinking. Probability, too, if regarded as something endowed with some kind of
objective existence, is no less a misleading misconception, an illusory attempt
to exteriorize or materialize our true probabilistic beliefs.
 \end{quote}

\medskip

But if entropy is supposed to be a numerical weight of a probability
distribution, and \emph{probability} does \emph{not} exist, then entropy is
indeed in big trouble!  A contemporary apologist\cite{Nau01}
tries to make  de Finetti's slogan more palatable by explaining,
 \begin{quote}
 [What he means is] that probability does not
exist in an \emph{objective} sense. Rather, probability exists only
\emph{subjectively} within the minds of individuals.
 \end{quote}
 
\noindent Good try! (I'm not being sarcastic.) The issue is, were we just
looking for probability in the \emph{wrong material object}---say, this
six-faced ivory die instead of the gambler's brain? No! The problem is not
\emph{where} we were looking, but \emph{what} we were looking for. 

Probability is not a \emph{fact}, but a \emph{working hypothesis}---the major
premise in a syllogism.

\medskip

Plato having said ``If (\emph{major premise}) all men are immortal---and
(\emph{minor premise}) Socrates is a man---then (\emph{conclusion})] Socrates
is immortal,'' on Socrates' death his wife Xantippe can tell Plato, ``I know
for a fact that Socrates was a man---your minor premise was right. But Socrates
is dead now---your major premise must have been wrong!''

In other words, Plato didn't dictate ``All men are immortal \dir{full stop} and
therefore Socrates (who's a man) is immortal \dir{full stop},'' but just
proffered a noncommittal ``\emph{If \dots then.}''

\medskip

Jaynes paid a tribute to physics when he called his lifetime
project\cite{Jaynes03} {\sl Probability Theory: The Logic of Science}. He had
previously written a paper that carries a more \emph{general} title,
``Probability theory as logic''\cite{Jaynes90probAsLogic}. But, within this
more ecumenic framework, what distinguishes probability from ordinary logic? I
would have added ``Probability theory as logic \emph{in parallel}.''

\medskip

The idea is that the ``probability of an event'' is a specialized kind (not
just a vague analog) of a syllogism machine---let's call it
$\text{ParSyll}$---which takes as arguments a \emph{probability distribution}
$P$ for the major premise and an \emph{event} $x$ for the minor premise, and
returns as a conclusion a \emph{number} $p(x)$. Thus,
$p(x)=\text{ParSyll}(P,x)$.

\smallskip

The internals of this machine work as follows. The major premise $P$ is a
lookup table---a list of primitive objects called \emph{outcomes} each
accompanied by a numerical \emph{weight}. The minor premise $x$---an
\emph{event}---is any collection of outcomes---typically specified by some
criterion related to the problem at hand. Given the set $x$, ParSyll breaks it
up into its elements and looks up in $P$ each of these elements separately and
independently, adding up the resulting weights as they come. Since the order of
the addends is irrelevant, the lookups (each of which is in essence a
``microsyllogisms'') can be performed in any order or concurrently, and thus
``in parallel.'' The total accumulated weight is output as a conclusion---the
\emph{probability of that event}.

\xtra{A nurse is filling out a form for a patient who happens to be an
\emph{event}: ``Eye color?'' ``Brown!'' ``Height?'' ``Five foot nine!''
``Probability?''  ``How would I know?'' ``You're right, I have to look it up in
\emph{my database}. One moment \dots\ today your probability is 1/4!'' Note
the different \emph{direction} of information flow for eye color and
probability.}

\medskip Jaynes calls ``mind projection fallacy'' the ``confusion between
reality and a state of knowledge about
reality.''\cite{Jaynes90probAsLogic,Jaynes03} ``The belief that probabilities
are realities existing in Nature is pure mind projection fallacy,'' since a
probability ``is something that we \emph{assign}, in order to represent a state
of knowledge,'' while counts and frequencies are ``factual properties of the
real world that we \emph{measure} or \emph{estimate}.'' [emphasis mine]

This confusion is certainly facilitated when a verbal \emph{shortcut}---such as
using ``my'' in different senses (\cf the climax ``my nose, my dog, my son, my
wife, my job, my country, my God)---crystallizes into a mental groove.

\xtra{The ``ambition prize'' for \emph{mental projection fallacy} (the
philosophical term for this is \emph{reification}, that is, ``turning a mental
construct into a \emph{thing}'') must be given to Anselm of Canterbury's
ontological proof of God's existence (1078): (\emph{major premise}) God is a
being than which nothing greater can be conceived; (\emph{minor premise}) One
can conceive of a God $A$ existing in one's mind, even if one denies the
existence of this God $A$ in reality; (\emph{reduction ad absurdum}) Then one
can conceive of a being $B$ just like God $A$ but having the additional
property of existing in reality; but (\emph{in contradiction to the minor
premise}) that being $B$ would be \emph{even greater} than God $A$,
(\emph{conclusion}) Therefore, God exists!

\cue{a colleague}{And, if I may ask, where did get your major premise?}}

\smallskip

The answer to the above question, or the equivalent ``And where did you get
your probability distribution?'' can only be ``I made it up, to put into
tentative use and see how well it works. ``What do you mean, `you made it up';
didn't you use the best of your knowledge to fabricate it?''  ``Not the best of
my \emph{knowledge}---that might be the case, but is irrelevant. I put into it
some specially crafted, pointed working hypotheses, to see where they would
lead.  Or maybe I put into it a colleague and competitor's presented data
(about which I'm suspicious), to see whether they would lead to unbelievable
conclusions.''
 If my probability distribution works well, whether to make earnest predictions
or to expose a colleague's fraud, I may encourage others to use it.'' This is
the gist, and the gift, of MaxEnt.

\medskip

It is well known that the exercise of logic never \emph{adds} to our knowledge:
its role is to make a certain aspect of that knowledge clearer or more
explicit, while keeping all the rest conveniently out of our sight.

If the machinery of probability is just ``logic in parallel,'' then using it
will never give us more than is contained in the probability distribution we
started with. So the question arises, ``When (in order to use the convenience
of that machinery) we turned our initial description in terms of more-or-less
vague constrains into a probability distribution---something more handy to
circulate and to play with---are we sure that we didn't throw out \emph{at
least some of} the baby with the bathwater?''

To skirt that dilemma, would it be possible, at least in principle, to skip the
probability stage? To directly go from a description to a recommendation for
action (say, to bet on a certain event), or to directly compute from a
description of the initial state our best description of the final state?
(Note that I leave open the possibility that such descriptions might not even
be expressible as probability distributions.)

\Subsect[nondeterministic]{Nondeterministic systems}

With a \emph{one-to-many} (or \emph{nondeterministic}) law, a trajectory may
fork out into many branches. We have explicitly dealt with many-to-one laws, as
in the Day 2 dialog and in \sect{2nd}, but not with \emph{one-to-many} laws. In
the two dialogs of Day 3 (\sect{increase}), we ascribed the de facto
nondeterminism to external ``perturbations,'' not to the internal law.

One-to-many laws of a genuine internal nature are encountered, for instance, in
automata theory. In physics and many other fields, on the other hand,
one-to-many laws usually arise from an attempt to incorporate (for theoretical
or practical convenience) into the \emph{internal} laws of a nominally
self-contained system the effects of \emph{small pertubations} of an
\emph{external nature}, especially when the latter are statistically
predictable---as in Boltzmann's well-known \emph{gas transport equation}.

We could have easily added a third column to table \eq{budget} to account for
formally \emph{nondeterministic} dynamical systems. However, while its
contribution to the top row (internal entropy budget) would have been
``$\geq0$,'' and thus different from the ``$=0$'' of the second column
(invertible), the {\sc total} would have been the \emph{same} for both columns.
 Thus a third column for one-to-many laws wouldn't have introduced a very
informative distinction: internal nondeterminism and external noise are hard to
tease apart in theory as well as in practice.

\medskip

In physics, this ambiguous aspect of nondeterminism is always lurking in the
background, when it is not stomping in the foregroud. All this, while exciting
in itself, is only marginally relevant to the present exploration of the ``core
meaning'' of entropy, and we shall not delve into it here.

 \iftrue When Einstein said ``God doesn't play dice'' he didn't close his eyes
to certain nondeterministic aspects of quantum mechanics. But he wondered
whether quantum mechanics was a \emph{complete} description, and whether a more
complete description would have done away, at least in principle, with that
nondeterministic component. After all, the randomness of a die's throw can
easily be explained as a form of \emph{deterministic chaos}!

The problem with the latter suggestion is that, as long as our description of
the state of the universe is given by a conventional \emph{probability
distribution}, such an explanation by itself will not account for the violation
of Bell's inequality displayed by certain statistical relations of quantum
mechanics; for that, we need a \emph{density operator} (\sect{entropy000}, von
Neumann).

The general consensus is that the fundamental dynamical laws of quantum
mechanics obey, much as those of classical mechanics, a \emph{strictly
one-to-one} paradigm. As for the \emph{interpretation} of some of its
nondeterministic aspects, it seems that, even after a century of deliberations,
the jury hasn't still made up a single mind.

For example, even though an ordinary probability distribution is viewed as the
default way to express one's prior knowledge about the state of a system, the
violation of Bell's inequality by certain quantum-mechanical effects seems to
rule out the use of a mere \emph{probability distribution} as a consistent
\emph{prior} for the universe as a whole.

In his irreverent romp\cite{Aaronson13QC}, Scott Aaaronson's says
 \begin{quotation}
 Quantum mechanics is a beautiful generalization of the laws of probability, a
generalization based on the 2-norm rather than the 1-norm, and on complex
numbers rather than nonnegative real numbers.
 \end{quotation}

\noindent These issues (prefigured by the replacement of a \emph{probability
distribution} by a \emph{density operator}; see \sect{entropy000}, \emph{von
Neumann entry}) are exciting in themselves, but only marginally relevant to the
present exploration of the ``core meaning'' of entropy, and we shall not delve
into them here.\fi

 \Subsect[many]{The ``many-scenarios interpretation'' of probability}

A the end of \sect{prob} we actually raised two questions, namely,
 \begin{enumerate}
 \item Let us go (say, via MaxEnt) from a description---in terms of facts
stated or constraints given---to a probability distribution. Is such a
conversion \emph{information lossless}?
 \item Doesn't a description in terms of facts stated or constraints given
constitute by itself a \emph{bona fide} major premise? And doesn't the passage
of time for a dynamical system, together with a list of the random disturbances
affecting its internal dynamics as well as the degradations externally
introduced by our abridged bookkeeping---doesn't all of this constitute by
itself a minor premise?  If so, can't a plain---though possibly
gigantic---syllogism derive the desired conclusion from those major and minor
premises?
 \end{enumerate}

Before tackling the above two questions, let's dwell first on a difficulty that
affects both. Namely, many descriptions are too vague to be usable \emph{by
themselves} as a major premise or as a recipe for a probability
distribution. For example, if you intend to use the MaxEnt principle, what
range of probability distributions may satisfy a constraint such as ``Tomorrow
is going to be much warmer?''  Approaches like Zadeh's \emph{fuzzy
logic}\cite{Zadeh65fuzzy}, that quantize warmness as, say, lukewarm, warmish,
warm, very warm, and hot, haven't made inroads in spite of their superficial
appeal---and not much else seems to be on the horizon.

\medskip

I will dispose summarily of Question 1, unfortunately because at the moment I'm
not quite sure of the answer. What first comes to my mind is, ``Can one go back
from a probability distribution to the set of constraints that generated it via
MaxEnt? (Or at least to an acceptable \emph{equivalence class} of
constraints?)''  If one could do that in a systematic way, then the two
formulations would be equivalent, and the transformation from one and the other
(and vice versa) would be invertible and thus information-lossless.

In this repect, let me quote Jaynes\cite{Jaynes93backward} (the emphasis is
mine):
 \begin{quote}
 Fuzzy Sets are (quite obviously, to anyone trained in Bayesian inference)
crude approximations to Bayesian prior probabilities. They were created only
because their practitioners persisted in thinking of probability in terms of a
``randomness'' supposed to exist in Nature but never well defined; and so
concluded that probability theory is not applicable to such problems. \emph{As
soon as one recognizes probability as the general way to specify incomplete
information}, the reason for introducing Fuzzy Sets disappears.
 \end{quote}
 The ambiguity here is in that ``specify incomplete information'' clause.  Does
it mean, as Jaynes seems to indicate, ``completely capture'' the incomplete
information, or simply ``do the best we can with the given information, even
though we have no way to use all of it?'' In the ``much warmer'' example given
above, it seems to me that all that one can consistently use is the constraint
``greater than,'' while the ``much'' qualifier, without further qualification
or some usage statistics, must unfortunately be discarded. And that even though
we feel that ``much warmer'' somehow conveys more information than just
``warmer.''

\medskip

Coming to Question 2, just as we started---\`a la Boltzmann---with entropy as a
\emph{count}, now let us start---\`a la Laplace---with probability as a
\emph{ratio} of counts, \ie favorable over total. Here there is no need to
externally assign different weights to different outcomes via a probability
distribution, since the weight of an outcome is a naturally defined internal
quantity---simply the number of representative points---the ``favorable
count''---of that outcome. The outcomes thus make up a \emph{multiset}, and the
events are all the possible \emph{subsets} of outcomes.

\xtra{A \emph{multiset} is like an ordinary set, except that any element may
appear in it in \emph{multiple instances} (rather than a single one). Let $X$
be the collection of all these ``sets of clones'' or \emph{outcomes}.  If
$N=\sum_{x\in X}n_x$ is the total number of elements in the multiset (counting
each outcome $x$ with its multiplicity $n_x$), then the collection
$\{n_x/N\,|\,x\in X\}$ is evidently a probability distribution.}

In the simplest case, with just two outcomes, 0 (for ``heads'') and 1 (for
``tails''), the four possible events $\{\}$, $\{0\}$, $\{1\}$, and
$\{0,1\}$---representing respectively ``neither heads nor tails,'' ``heads,''
``tails,'' and ``heads or tails''---have counts of 0, 1, 1, and 2, which,
divided by the number 2 of outcomes, yield ``probabilities'' 0, $\frac12$,
$\frac12$, and 1. However, this model of an abstract ideal coin becomes
instantly useless as soon as we want to use it for a real coin---no matter
how ever-so-slight its bias may be. As they say, this model does not ``degrade
gracefully.''

Let us take a more flexible approach. Consider the collection {\bf P} of all
Python programs consisting of no more than two hundred characters and such
that, at a press of a button, each of these program will output (say, within a
thousand machine cycles) a single binary digit. Let the programs in this
well-defined collection be sequentially numbered $1,2,\dots,N$, where of course
$N$ is a hyper-astronomical number.  Let us run all these programs in parallel,
and collect into one set the $N$ binary digits generated by one button press,
each digit tagged by the number $n$ of the program it came from. Approximately
half of those digits will be 0's and the other half 1's. Now just ignore the
tags, and you'll have a \emph{multiset} with just two outcomes, 0 and 1, one
being present in $n_0$ ($\approx N/2$) copies and the other in $n_1$
($=(N-n_0)/2$) copies. We offer this deterministic piece of machinery as a
model of a (non-deterministic) \emph{coin-tossing machine}. To model coins with
different biases, just tinker with this collection {\bf P} of program by
throwing out of it a number of its elements. \emph{Unless you know well what
you are doing, you'll have a hard time getting out of the decimated collection
an expected value other than one close to some special value such as 0, 1/2, or
1}---just as with an \emph{ordinary} coin (have you \emph{tried} to machine a
coin that will reliably give heads 3.14\% of the time?). Of course, by running
this astronomical number of programs \emph{one-by-one} and selectively
discarding a fraction of them according to whether their output is 0 or 1, you
can ``easily'' make this coin tosser come up with any probability in
between---or behave deterministically for you while still remaining random for
the noninitiated. Physical situations of this kind are wonderfully discussed in
Jaynes's book\cite{Jaynes03}.

\smallskip

Going back to the statement of Question 2, that gigantic syllogism's
\emph{conclusion} may well turn out to be a \emph{gigantic table} with 0 and 1
values attached without apparent rhyme or reason to the zillion
\emph{individual outcomes} or ur-states that make up the event of concern---no
probability distribution was involved!  Then it is up to us to take the
responsibility---as a single optional final step---to summarize that table by
counting the 1's for each event and using those numbers as the \emph{weights}
of the respective events. 

That is no worse, and possibly better, than going through the
probability-distribution bottleneck, since in that way we keep \emph{all} of
the information we have till the end (that is why this approach is so
cumbersome) and then only at the last step throw away what we sensibly can. In
the traditional approach, we jettison most of what seems irrelevant (or just
too unwieldly) \emph{before} take-off; that makes for a lighter flight, but
will it take us to the same destination?

\medskip

If you are worried that this ``many-worlds'' approach may still not yield
enough randomness for the coin, I can throw in more layers of ``super-worlds''
(this is a mental experiment, after all) each having different rules for
generating its own lower-level many worlds, and so forth \emph{ad infinitum}. I
suspect that the limit of this process, which still uses purely
\emph{deterministic} resources at every level is the only definition of
\emph{randomness} that a Laplace would have been happy with.

 \Sect[dynamics]{The dynamics of entropy}

When we say ``the entropy of a black hole'' or ``entropy of the market'' we
must mean something like ``the entropy of (a probability distribution of
(quantum states of (\dots of (\dots of (\dots of (a certain kind of (black
hole))\dots))))). All of the ``of's'' in this hierarchy are like different
levels of civil servants with technical roles that are useful but not
necessarily very transparent. Moreover entropy, as a function that maps a
descritpion to a count, has no initiative---it ``just obeys orders!''  A
policy-maker must do a bit of homework up and down the hierarchy to figure out
the precise effect of an order given from the top of the hierarchical chain.

\xtra{Chains of command of this kind beginning with ``entropy'' are employed in
all sorts of disciplines. Some are short and widely used, and have developed
standards of usage and an agreed upon lingo.  So, when we say ``The entropy of
a Valentine message is less than a bit per typed character'' (think of the
endless repetition of `love' and `dove' and `blue' and `you'), everyone in the
data transmission business will understand that this entropy is not about a
single character, word, or message, but, skipping several levels of the
hierarchy, is about the \emph{language}---the vocabulary, the style, the
mannerism---used by Valentine chatter \emph{as a whole}.}

\medskip

Chains headed by ``entropy'' are less transparent in other situations. In
addition, there may be many chains of the form ``entropy of \dots $\langle$
something$\rangle$'' where the last link, $\langle$something$\rangle$, is the
same for both, but the internal makeup of the chain is different. So asking
``Mack, what's the entropy of that brick?'' may get you surprisingly different
responses even though they all have ``Mack'' and ``brick'' as first and last
link.

\smallskip

The problem all started with Clausius, who, as we've seen in \sect{entropy000},
was careful to coin for the quantity he had discovered a new-greek neologism
that could hardly be used for anything else. When he spoke of the ``entropy of
a transformation'' he had in mind not a function but some sort of fluid that
had coursed in or out of a certain object subjected to a certain treatment or
``transformation''---or even had been \emph{generated within it} in the course
of the treatment. In that context, there was only one thing that could be meant
by ``a certain amount of entropy.''

Luckily for Clausius's fame, entropy was eventually found to be not a physical
substance but an abstract concept of \emph{enormous generality}---a
\emph{counting} concept (\cf the \emph{Boltzmann} entry in \sect{entropy000}
and the Shepherd's Parable in \sect{not}). Consequently, today the kinds of
things that have a meaningful entropy associated with them are legion, and a
\emph{single thing} may have associated with it \emph{several
entropies}---which quantify different aspects of it or different levels of
description.

\bigskip

Certain of these new uses of entropy operate at a more abstract level than
Clausius's original construct. But they still play essentially the very
function which Clausius originally envisaged for his entropy, which is as
fundamentally important today as it was then. Namely, to determine whether
there are states of a dynamical system that are \emph{so easy to reach} from a
given state that the transition from the latter to one of them can occur just
about \emph{spontaneously}, and whether there are states that are instead
virtually \emph{unreachable}. Similarly, to determine under what conditions a
chemical reaction will run in one direction rather than the other, and what can
be done to make this reaction faster and more efficient, or, on the contrary,
to inhibit it.  All that is of ultimate importance in chemistry, biology,
materials science, energy conversion, and so forth.  Expectedly---but
regrettably---it is in this area that much confusion arises.

\smallskip

While the objective in those areas is to determine, at least in a qualitative
way, the \emph{dynamics} of a system by discovering which state transitions are
more likely, the strategy to calculate this has some affinity with a
\emph{static} variational method.  That is, (a) imagine a number of possible
states and the transitions between any two of these states, (b) compute the
entropy difference attached to each of these transitions, and then (c) consider
most likely those transition that are accompanied by the largest entropy
increases. All of this may sound like wordplay unless one understands the
underlying linguistic ambiguity. When people speak of a certain system
\emph{state} here, they actually mean a \emph{broad description}, and a system
is (tautologically) more likely to be in an ur-state matching a certain
description if this description is broader, that is, encompasses a \emph{larger
number} of ur-states.

\medskip

Naively, one may imagine that a transition from one description to another
reflects in some way an actual transition of the system from an \emph{ur-state}
to another according to the system's internal dynamics---but this need not be
the case! We have hinted at this at the end of \sect{entropy001}. Here, we
shall illustrate this in more detail by  a parable:

\xtra{A precious gold coin has been stolen from me; through an anonymous phone
call I learn that the coin may have been hidden in one of numerous sacs of
base-metal coins stored in a customs warehouse. I go there and ask to go
through the contents of the sacs to retrieve my gold coin. Given the vagueness
of the hint and the scale of the requested operation, to the customs officer
this seems a frivolous request, but he eventually allows me to go through one
\emph{single} sac of my choice. I should mark that sac now, but come back in
the morning for the actual search, since he wants to have with him a witness to
the operation. The sacs come in all different sizes. Not having any idea of
which sac my coin might be in, I of course choose the largest.

To better hide the coin, unbeknownst to me and the officer, the thief, who has
reasons to fear curious eyes, had been randomly moving it every night from one
sac to another.  I come in in the morning; officer and witness are already
there. Before I open the marked sac---call it sac $A$---a warehouse employee
remarks that a pile of several more sacs of coins had been found in a
corner. One of them---call it sac $B$---is much larger than sac $A$; I
ask---and am allowed---to switch my choice to sac $B$. I riffle through the
sac. Unsurprisingly, given the number of sacs, the coin is not there after all.

\smallskip

Now, I shall call ``ur-state'' of a coin (``microscopic'' state) the exact
physical spot where the coin finds itself, and ``state'' of the coin
(``macroscopic'' state) the \emph{sac} in which the coin is contained (as far
as I'm concerned, that level of detail is all that that matters).  Based on the
information I had at the moment, the \emph{most likely} state for the coin to
be in, the evening before, was sac $A$; by midmorning, it had moved to $B$; and
by the end of the morning, it had moved back to $A$ or to some other sac in the
newly dicovered pile (why?). As for the coin's ur-state, it had been shifted
during the night from a position inside some sac $C$ to a position inside a sac
other than $B$ (how do we know?).  That's it!

There was indeed a correlation (an \emph{implication}, in our case) between the
coin's \emph{internal} dynamics---that of its \emph{ur-state}---and that of its
\emph{state}: the coin's \emph{micro}location implies what sac it is in, while
the converse is not true---being in a certain sac says nothing about
\emph{where} in the sac. On the other hand, in our scenario the correlation
between the internal dynamics and that of my ``best-choice'' sac (based merely
on sac size) is very shallow: when this choice moved from sac $X$ to sac $Y$,
that didn't mean that the coin itself had moved from a position inside $X$ to
one inside $Y$. In our contrived scenario, my best-choice sac changed quite
independently of the movements of the coin itself!}

\bigskip

In the combinatorics of large systems, there is often a much better correlation
than this between microscopic and macroscopic dynamics. (More honestly said, we
don't even take into consideration a proposed macroscopic dynamics unless it is
strongly implied by the microscopics---When it's dark, we look for the key only
near a lampost; where else?)

Thus, in 100 coin tosses, the expected number of heads is 50, and even though
it is in priciple possible to get as a result the ur-state ``0 heads,'' a state
with $50\pm8$ heads is indeed \emph{overwhelmingly} more likely. In the above
gold coin example, if one sac was way larger than the aggregate of all the
others, for that very ``reason''(?) the coin would be virtually certain to
contain it.  You will agree that this ``reason''---the argumentation used by
entropy---is of a very different kind than ``the coin is likely to be there
because I put it there.''

\smallskip

You might conclude from all this that thermodynamics is but a collection of
trivial tautologies---and you'd of course be right. But \emph{precisely that} is
what makes the laws of thermodynamics so universal and so useful.

\Sect[conclusions]{Conclusions}

Returning to the definition of \emph{honest entropy} in \sect{contract} and its
illustration by an executive/consultant/auditor contract, we've seen that the
entropy ``of a system'' is  that of this continually adjusted probability
distribution representing the state \emph{of a model}. Every act of
approximation, truncation, compression, erasure, etc.\ that entails loss or
discarding of information gives a \emph{positive} contribution to the system's
\emph{honest} entropy. \emph{Negative} contributions could only come from
\emph{peeking} (which we have ruled out by contract; see \sect{ehren}), or from
losses of \emph{internal} entropy, which, as we've seen, can only arise from
the possible \emph{noninvertibility} of the internal dynamics.

\medskip

In a nutshell, if the consultant's deliverable is thought of as the
\emph{expected value} of a system variable, then this deliverable will be
deemed useless if not accompanied by a \emph{standard deviation} honestly
calculated according to an accepted discipline of \emph{propagation of
uncertainty}.  With an appropriate scaling, the latter is what we call
\emph{honest entropy}.

A consultant need to be kept honest in spite of the obvious lure of claiming
for his forecasts a smaller entropy burden than they actually bear.  To this
purpose the executive may occasionally use a more expensive \emph{auditor}
(peer review?) to go through the consultant's books and verify that the
latter's entropy was appropriately adjusted upwards every time he took a
shortcut.

\smallskip

If entropy is a way to quantify the ``known unknowns'' of a system's state,
usually representing the diffusive spread, the blurring, the unavoidable
\emph{random error} attendant to the use of a low-resolution model, what can
one say about estimating ``unknown unknowns,'' that is, the unintentional
\emph{systematic errors} that a model may introduce? Entropy is silent on this.
One may reasonably demand \emph{honest} accountants, but for the latter
estimate one would need \emph{imaginative}, \emph{out-of-the-box},
``\emph{creative}''---as it were---accountants; and that is an oxymoronic job
description that brings in drawbacks of its own!  So we'll leave that job to
the most competent specialists of each discipline---peer review, again?

\smallskip

 In  Shakespeare's memorable words,\Foot
 {{\sl Henry The Fourth, Part I} Act 3, scene 1, 52–-58.}

 \begin{quote}
 \begin{dialog}
    \cue{Glendower}{I can call spirits from the vasty deep.}
    \cue{Hotspur}{Why, so can I, or so can any man;
    but will they come when you do call for them?}
    \cue{Glendower}{Why, I can teach you, cousin, to command the devil.}
    \cue{Hotspur}{And I can teach thee, cousin, to shame the devil
    by telling the truth. Tell truth and shame the devil.}
 \end{dialog}
 \end{quote}

\bigskip

With the concept of \emph{honest entropy} I tried to capture a
physicist's---but not \emph{only} a physicist's---idealization of something
like a ``best accounting practice'' when making predictions.  If you can,
measure; if you can't measure, peek. If you can't either, all you can do is
imagine and compute---but, if you care about your reputation, attach to your
predictions their ``honest entropy'' margin of error.

\bigskip

If I had to explain entropy standing on one leg instead of writing a
thirty-page article, I would ask, ``Why is parking into a tight spot harder
than coming out of it? Note that the process is mechanically invertible---for
every path going in there is a well-defined time-reverse path coming out.''

 \xtra{{\sc answer:} (\emph{I'm dead serious}) Because of the second law of
thermodynamics---There are many more ways (greater entropy) for my car to be in
the middle of the street than in a tight parking spot!}

\medskip

If then I had to explain honest entropy standing on one big toe, I would
answer, with Joseph Stalin, ``It's not the votes that count, but who counts the
votes!''

\Sect[ack]{Acknowledgments}

I would like to thank Charles Bennett, Ed Fredkin, Lev Levitin, Norman
Margolus, and Sandu Popescu, who have been a lifelong source of inspiration,
encouragement, and criticism in the above matters. Special thanks to Gregg
Jaeger for his maieutic assistance in the conception and delivery of this
paper. Eiko Asazuma and Silvio Capobianco read the draft and helped improve it
by spotting blemishes and suggesting enhancements.

My warmest thanks to referees 1 and 2 for their touching encouragement and
solicitous and constructive criticism.

\end{document}